\documentclass[namedreferences,hyperref,optionalrh]{spr-sola}

\usepackage{graphicx}
\usepackage{booktabs}
\usepackage{amsmath}
\usepackage{url}
\usepackage{xcolor}

\newcommand{\arcsec}{^{\prime\prime}}

\makeatletter
\newcommand{\ion}[2]{#1\,\@Roman{#2}}
\makeatother

\chardef\us=`\_

\begin{document}

\begin{frontmatter}

\title{\ion{Mg}{2} Signatures of Frequency Dependent Wave Propagation in Magnetic Elements}

\author[addressref=aff1,corref,email={d.baker2060@gmail.com}]{\inits{D.}\fnm{Deborah}~\snm{Baker}\orcid{0000-0002-0665-2355}}
\author[addressref=aff2]{\inits{D.M.}\fnm{David~M.}~\snm{Long}\orcid{0000-0003-3137-0277}}
\author[addressref=aff3]{\inits{M.}\fnm{Marco}~\snm{Stangalini}\orcid{0000-0002-5365-7546}}
\author[addressref={aff4,aff5}]{\inits{M.}\fnm{Michele}~\snm{Berretti}\orcid{0009-0007-2465-1931}}
\author[addressref={aff6,aff7}]{\inits{A.S.H.}\fnm{Andy~S.~H.}~\snm{To}\orcid{0000-0003-0774-9084}}
\author[addressref={aff8,aff1,aff9}]{\inits{S.L.}\fnm{Stephanie~L.}~\snm{Yardley}\orcid{0000-0003-2802-4381}}
\author[addressref=aff10]{\inits{P.}\fnm{Paola}~\snm{Testa}\orcid{0000-0002-0405-0668}}

\address[id=aff1]{University College London, Mullard Space Science Laboratory, Holmbury St Mary, Dorking, Surrey, RH5 6NT, UK}
\address[id=aff2]{Centre for Astrophysics \& Relativity, School of Physical Sciences, Dublin City University, Glasnevin Campus, Dublin, D09 V209, Ireland}
\address[id=aff3]{ASI Italian Space Agency, Via del Politecnico, s.n.c., I-00133 Roma, Italy}
\address[id=aff4]{University of Trento, Via Calepina 14, 38122 Trento, Italy}
\address[id=aff5]{University of Rome Tor Vergata, Department of Physics, Via della Ricerca Scientifica 3, 00133 Rome, Italy}
\address[id=aff6]{ESTEC, European Space Agency, Keplerlaan 1, PO Box 299, NL-2200 AG Noordwijk, The Netherlands}
\address[id=aff7]{Universit\'e Paris-Saclay, CNRS, Institut d'Astrophysique Spatiale, Orsay, France}
\address[id=aff8]{Department of Mathematics, Physics and Electrical Engineering, Northumbria University, Newcastle Upon Tyne, NE1 8ST, UK}
\address[id=aff9]{Donostia International Physics Center (DIPC), Paseo Manuel de Lardizabal 4, San Sebasti{\'a}n, 20018, Spain}
\address[id=aff10]{Harvard-Smithsonian Center for Astrophysics, 60 Garden St, Cambridge, MA 02193, USA}

\runningauthor{Baker et al.}
\runningtitle{\ion{Mg}{2} Signatures of Frequency Dependent Wave Propagation}

\begin{abstract}
Waves in solar magnetic elements provide a pathway for transporting and dissipating energy in the chromosphere, yet the relationship between photospheric oscillations and chromospheric spectral structure remains incompletely understood. 
We investigate whether distinct chromospheric \ion{Mg}{2} spectral states are linked to different populations of photospheric kink mode oscillations, with particular focus on the role of high-frequency ($\sim$5\,mHz) waves. 
Magnetic elements are tracked in \textit{SDO}/HMI magnetograms to derive horizontal velocity power spectra, while corresponding \ion{Mg}{2} $h+k$ profiles from \textit{IRIS} are classified using $k$-means clustering into morphology-based (SHAPE) and intensity-based (ABS) groups. 
Both classifications exhibit systematic differences in oscillatory behavior. The SHAPE clusters isolate spectral morphologies associated with enhanced high-frequency power, including cases where oscillations in the $4$--$6$\,mHz band exceed those in the photospheric $p$-mode band. 
In contrast, the ABS clusters show a monotonic increase in the relative contribution of high-frequency oscillations with increasing emission strength and magnetic flux. 
These results indicate that high-frequency oscillations are both selectively enhanced in specific chromospheric states and systematically strengthened in stronger magnetic environments. 
This suggests that magnetic structuring plays a key role in regulating both the transmission and the conversion of kink mode wave energy into observable chromospheric signatures.
\end{abstract}


\end{frontmatter}

\section{Introduction} \label{sec:intro}

Waves are continuously generated in the solar photosphere by convective motions, both directly and through the excitation of global $p$‑mode oscillations.
During propagation through the magnetized atmosphere, wave behavior is governed by the coupled plasma and magnetic field system, which can in turn be modified by the waves themselves.
The photospheric acoustic power is concentrated near 3\,mHz, reflecting the dominant contribution from global $p$-mode oscillations.
At chromospheric heights, however, enhanced acoustic power is frequently observed near 5\,mHz \citep[e.g.][]{Howard1968,Lites1986,Khomenko2015,Jess2015}. 
This shift toward higher frequencies has often been attributed to atmospheric filtering effects, whereby waves with frequencies above the local acoustic cutoff are able to propagate, while lower-frequency waves become evanescent. 
The effective cutoff can be modified by factors such as magnetic field inclination and radiative losses, allowing lower-frequency waves to propagate to higher layers of the solar atmosphere under certain conditions \citep[e.g.][]{Centeno2006,Jefferies2006,Wedemeyer2009,Jimenez2011}.

Recent observations nonetheless suggest that this picture may be incomplete and, in addition to magneto-acoustic waves, several MHD resonant modes can be excited at the same time in magnetic tubes.  
While small scale photospheric magnetic elements are expected to be driven primarily by granular forcing and the ambient $p$-mode oscillations, \cite{Berretti2024} found that their horizontal oscillations exhibit a pronounced peak near 5\,mHz. 
These oscillations are interpreted as kink modes of magnetic flux concentrations, with a characteristic frequency more commonly associated with chromospheric oscillations than with a photospheric driver.
The origin of this behavior remains unclear, but it indicates that magnetic structures may preferentially channel or enhance oscillatory power near 5\,mHz already within the photosphere. 
Such kink oscillations are also thought to contribute significantly to the transport of energy into the upper solar atmosphere, where they may play an important role in atmospheric heating and solar wind acceleration \citep[e.g.][]{Tomczyk2007,Magyar2021,Gao2024}.

A key question is whether these horizontal kink oscillations leave a detectable imprint on the chromosphere. 
If wave energy is efficiently transported through magnetic structures at these frequencies, it may influence the thermodynamic and spectroscopic properties of the overlying atmosphere.
Establishing such connections would help identify where wave energy is deposited and dissipated, while also providing insight into the origin of the pronounced 5\,mHz peak observed in photospheric kink oscillations.

\begin{figure}
\centering

\includegraphics[width=0.5\textwidth]{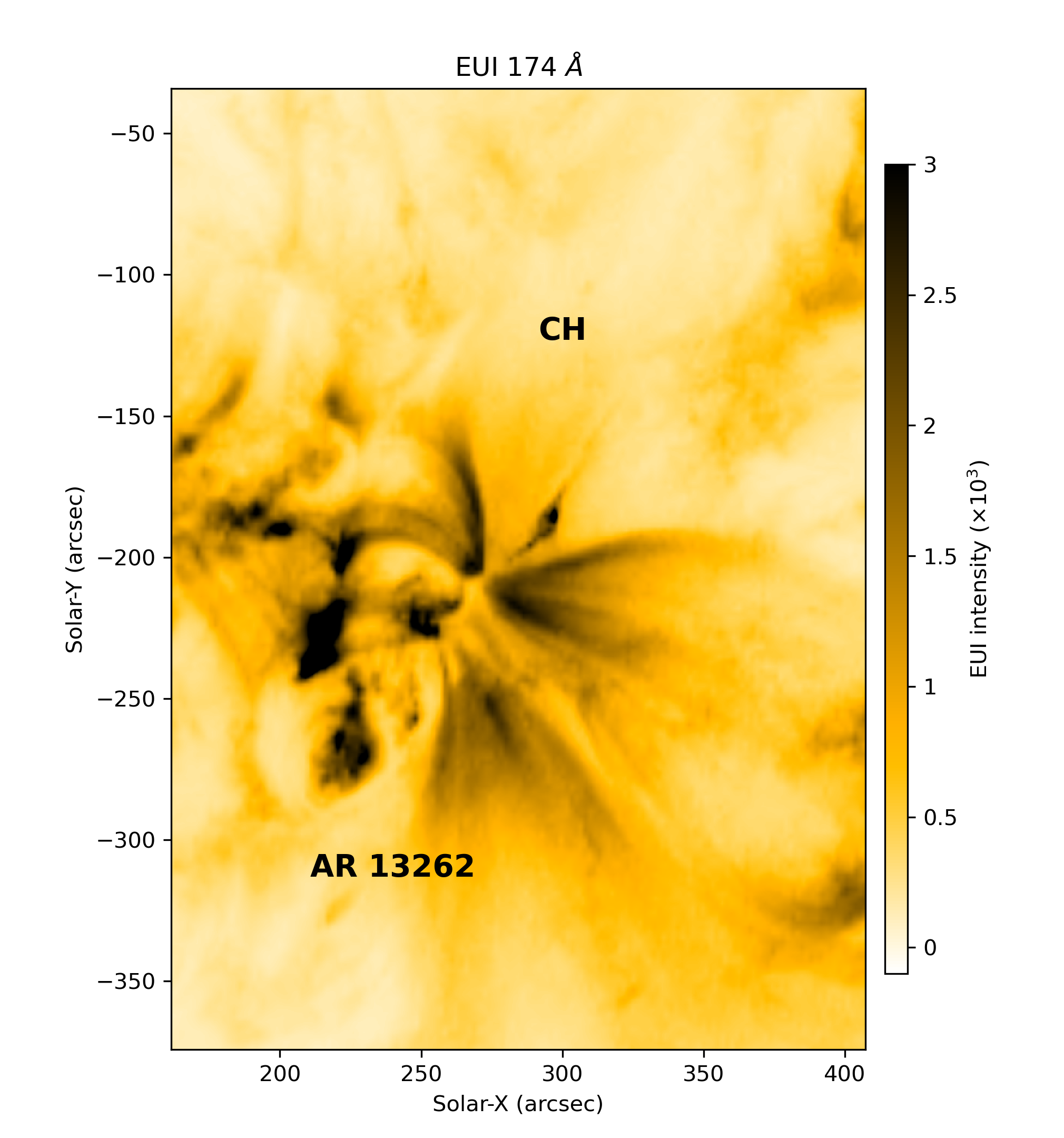}\hspace{-0.04\textwidth}
\includegraphics[width=0.5\textwidth]{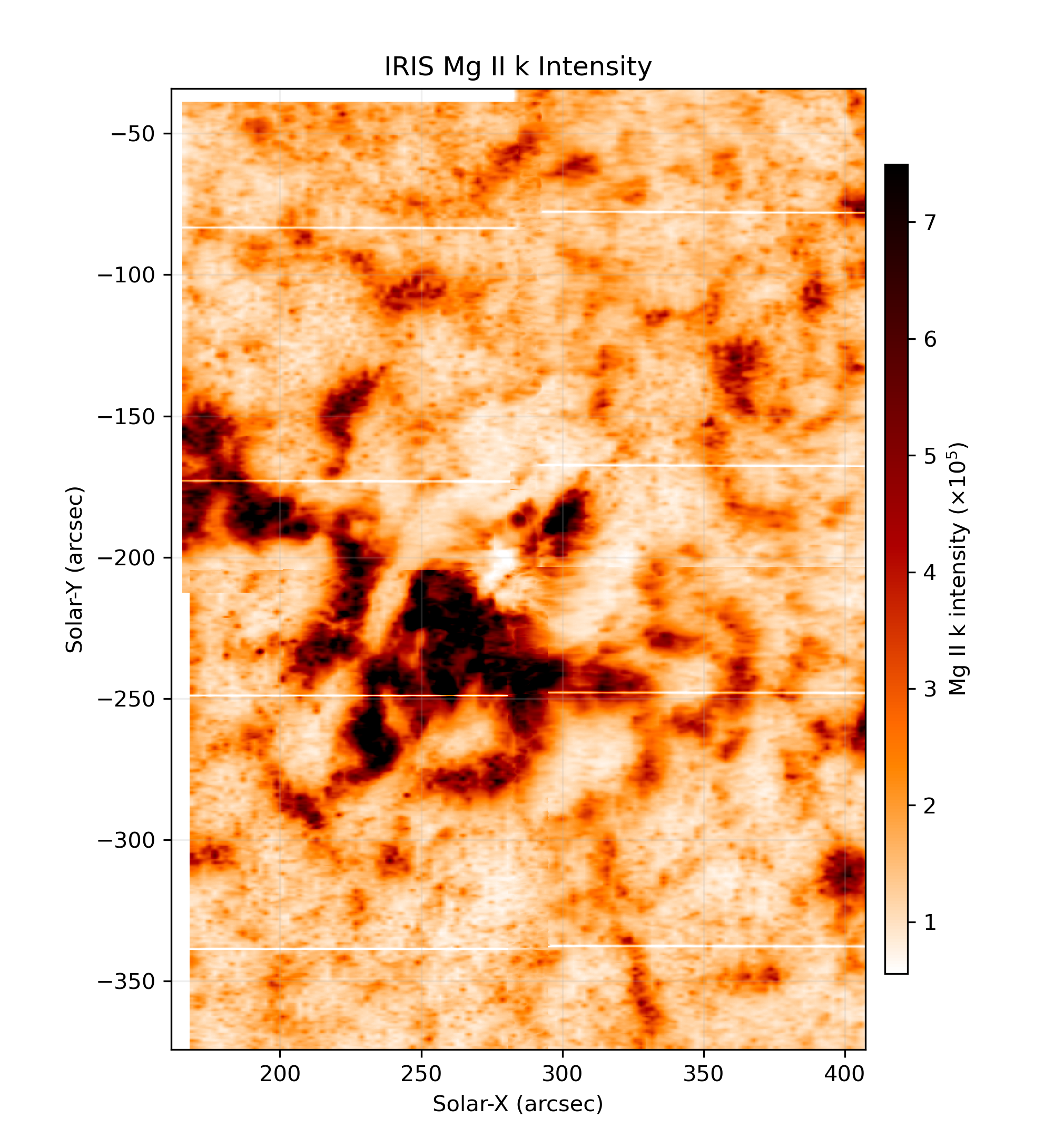}

\vspace{-0.45cm}

\includegraphics[width=0.5\textwidth]{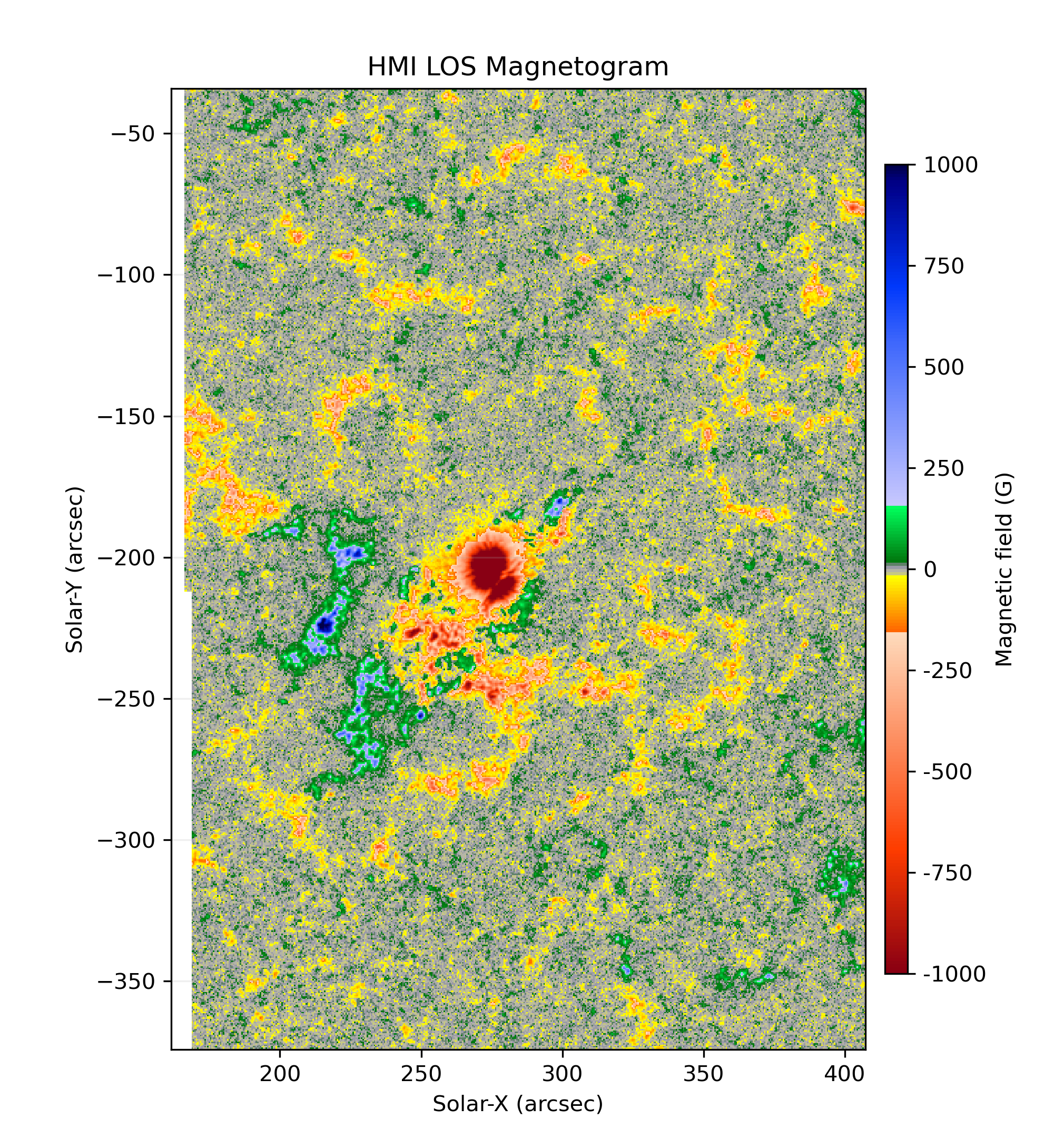}\hspace{-0.04\textwidth}
\includegraphics[width=0.5\textwidth]{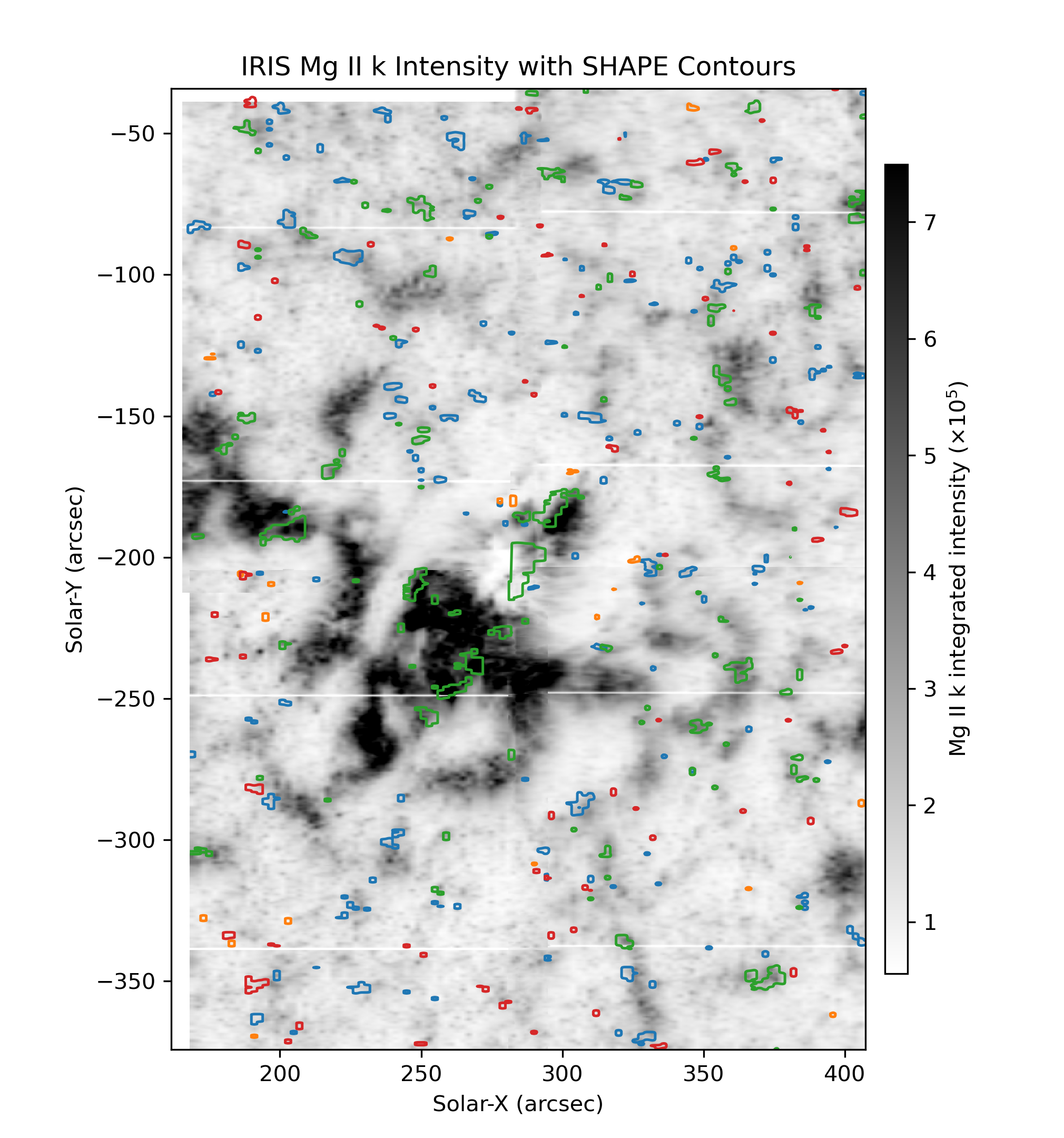}

\vspace{-0.4cm}
\vspace{0.15cm}
\caption{Active region--coronal hole complex on 2023 March 29. 
Top left: EUI 174\,\AA\ image at 18:14~UT. 
Top right: IRIS \ion{Mg}{2} $k$ 2796.35\,\AA\ intensity mosaic, composed of tiles beginning at 15:52 (upper right), 16:55 (upper left), 17:38 (lower right), and 18:44~UT (lower left). 
Bottom left: HMI LOS magnetogram mosaic saturated at $\pm100$~G.
Bottom right: IRIS \ion{Mg}{2} $k$ 2796.35\,\AA\ intensity mosaic with contours identifying magnetic elements color-coded with SHAPE centroids in Figure~\ref{fig:fig2} (SHAPE 1--4 green/blue/red/orange).
All intensity images are displayed using reversed color tables.}
\label{fig:fig1}
\end{figure}

Among the most powerful diagnostics for investigating wave processes in the chromosphere are the \ion{Mg}{2} $h+k$ resonance lines observed with the \textit{Interface Region Imaging Spectrograph} \citep[IRIS;][]{Depontieu2014}. 
Formed under conditions of strong scattering and partial frequency redistribution, these lines sample atmospheric layers extending from the upper photosphere to the upper chromosphere and lower transition region \citep{Leenaarts2013a,Leenaarts2013b,Pereira2015}. 
The far wings originate low in the atmosphere, while the $k_2$ and $h_2$ emission peaks and the central reversals $k_3$ and $h_3$ form progressively higher in the chromosphere. 
As a result, \ion{Mg}{2} profiles provide a sensitive diagnostic of chromospheric structure and dynamics over a broad range of heights.
Their extended formation range makes them particularly well suited for investigating how wave energy and magnetic structuring influence chromospheric conditions \citep{Schmit2015,DePontieu2021}.

To explore these relationships, we analyze the chromospheric spectral properties of an active region–coronal hole (AR–CH) complex in relation to the kink mode oscillations detected in photospheric magnetic elements.
Across the mosaic field of view (FOV), the identified magnetic elements sample a broad range of magnetic environments.
The IRIS \ion{Mg}{2} $h+k$ spectra corresponding to each structure are classified using unsupervised $k$-means clustering, enabling an objective identification of recurring chromospheric spectral classes across the diverse environments present within the AR–CH complex.
The oscillatory properties of the corresponding magnetic elements are then characterized through their power spectra, allowing the wave-power distributions associated with each spectral class to be directly compared.

This approach enables us to address a central question: do distinct chromospheric spectral states correspond to different populations of kink mode oscillations?
In particular, we examine whether specific spectral classes preferentially host power concentrated near the photospheric $p$-mode band ($\sim$3\,mHz) or at higher frequencies near and above $\sim$5\,mHz. 
By linking chromospheric spectral structure with the frequency content of photospheric kink oscillations, we seek to identify the atmospheric conditions that favor the propagation, amplification, and dissipation of different wave populations.

\begin{table}[!t]
\centering
\caption{Summary of the IRIS mosaic observations.}
\label{tab:IRIS}
\renewcommand{\arraystretch}{1.15}
\begin{tabular*}{0.90\textwidth}{@{\extracolsep{\fill}}ccc@{}}
\toprule
\multicolumn{3}{c}{\textbf{IRIS mosaic raster positions on 29 March 2023}} \\
\midrule
Time (UT) & Program Description & Raster Center (Solar-X, Solar-Y) \\
\midrule
15:52--16:25 & Orbiter RS Burst Mosaic Part 1 & $(321'', -119'')$ \\
16:55--17:29 & Orbiter RS Burst Mosaic Part 2 & $(213'', -120'')$ \\
17:39--18:12 & Orbiter RS Burst Mosaic Part 3 & $(337'', -285'')$ \\
18:44--19:18 & Orbiter RS Burst Mosaic Part 4 & $(232'', -286'')$ \\
\bottomrule
\end{tabular*}
\end{table}

\section{Data and Measurements} \label{sec:data}
\subsection{IRIS \ion{Mg}{2} Observations}\label{IRIS_obs}
Observations from IRIS were obtained on 29 March 2023 as part of a coordinated observing campaign with Solar Orbiter \citep{Muller2020}. 
The IRIS program (OBS 3400113362) employed a 64-step dense raster with a step size of 1.995$\arcsec$ and an exposure time of 30\,s at each step, producing a FOV of $\sim$ 127$\arcsec \times$ 175$\arcsec$ with a raster cadence of approximately 34 min. 
Spatial offsets between successive pointings extended the coverage in both the solar-X and solar-Y directions, resulting in a mosaic of four tiles with a FOV of $\sim$ 250$\arcsec$ in solar-X and $\sim$ 350$\arcsec$ in solar-Y (see Figure \ref{fig:fig1}).

The broad range of magnetic environments, including the large sunspot of AR 13262, surrounding plage, dispersed network fields, and the adjacent CH, sample a wide range of sizes, magnetic fluxes, and local magnetic conditions. 
Such diversity provides an excellent opportunity to investigate how chromospheric spectral properties and photospheric oscillatory behavior vary across different magnetic environments.

Figure \ref{fig:fig1} (top panels) shows a reprojected Solar Orbiter/EUI \citep{Rochus2020} 174\,\AA\ image of the observed AR--CH complex together with the corresponding \ion{Mg}{2} $k$ 2796.35\,\AA\ integrated intensity mosaic. 
The EUI image provides coronal context for the region studied here. 
The lower-left panel shows the HMI LOS magnetogram, illustrating the photospheric magnetic structure underlying the chromospheric emission observed by IRIS. Table~\ref{tab:IRIS} summarizes the four raster positions that comprise the IRIS mosaic, including their observation times, program descriptions, and raster center coordinates.

\subsection{Detected Photospheric Horizontal Oscillations}\label{sec:osc_obs}
To characterize the oscillatory properties of photospheric magnetic elements observed throughout the IRIS mosaic, we apply the methodology of \citet{Berretti2024} to full-cadence LOS magnetograms from the Helioseismic and Magnetic Imager \citep[HMI;][]{Scherrer2012} aboard the Solar Dynamics Observatory \citep[SDO;][]{Pesnell2012}. 
Magnetic elements are detected and tracked using the Solar Feature Tracking (SoFT) algorithm \citep{Berretti2025a,Berretti2025b}. 
For each IRIS mosaic tile, 33 consecutive magnetograms are selected, spanning a 25\,min interval. 
The magnetograms are co-registered to remove solar rotation, ensuring a fixed FOV and enabling consistent tracking of magnetic elements through time.
Specifically, the HMI image time series was co-registered using a template-matching procedure based on normalized cross-correlation. 
For each tile, the first image was used as the reference template, and spatial shifts were determined by cross-correlating the template with subsequent frames. 
This allowed the time series to be co-aligned prior to feature tracking and velocity extraction.

SoFT identifies and follows magnetic structures using image segmentation combined with frame-to-frame feature association. 
Two detection modes are available: a \textit{fine} mode that resolves individual small scale magnetic elements and a \textit{coarse} mode that groups neighboring features into larger magnetic concentrations. The fine mode is adopted here because the objective is to measure the horizontal motions of individual magnetic elements across the full range of magnetic environments sampled by the mosaic. 
In this mode, the large sunspot of AR 13262 is partitioned into multiple features associated with local magnetic maxima. 
Changes in the morphology of the sunspot magnetic field can occasionally produce large apparent displacements of the barycenters of these features, resulting in unrealistically high horizontal oscillations. 
Features exhibiting such behavior are excluded from the subsequent analysis.

Detection begins by masking pixels below a threshold derived from the magnetogram noise level \citep[approximately $6$\,G;][]{Liu2012}. 
Local maxima are then identified with a minimum separation of three pixels to avoid fragmentation of individual structures. 
A distance-transform segmentation partitions the magnetic signal into distinct labeled regions, and structures smaller than four pixels are discarded to suppress spurious detections.

Tracking between frames is based on spatial overlap of the segmented regions. 
A feature in frame $n$ is linked to the feature in frame $n+1$ with the largest number of shared pixels, and the match is retained only when the association is reciprocal. 
This procedure allows magnetic elements to be followed through the full time series. 
Physical properties are then derived for each magnetic element, including positions determined from the magnetic signal-weighted barycenters of the pixels, magnetic field strengths, and magnetic fluxes.
Horizontal oscillations are obtained from the temporal evolution of the barycenter positions using a first-order time derivative. 
Power spectra are subsequently computed from the resulting velocity time series, providing a characterization of the horizontal (kink mode) oscillations of the magnetic elements \citep{Berretti2024}.

\begin{figure}
\centering
\begin{minipage}{0.48\textwidth}
\centering
\includegraphics[width=\linewidth]{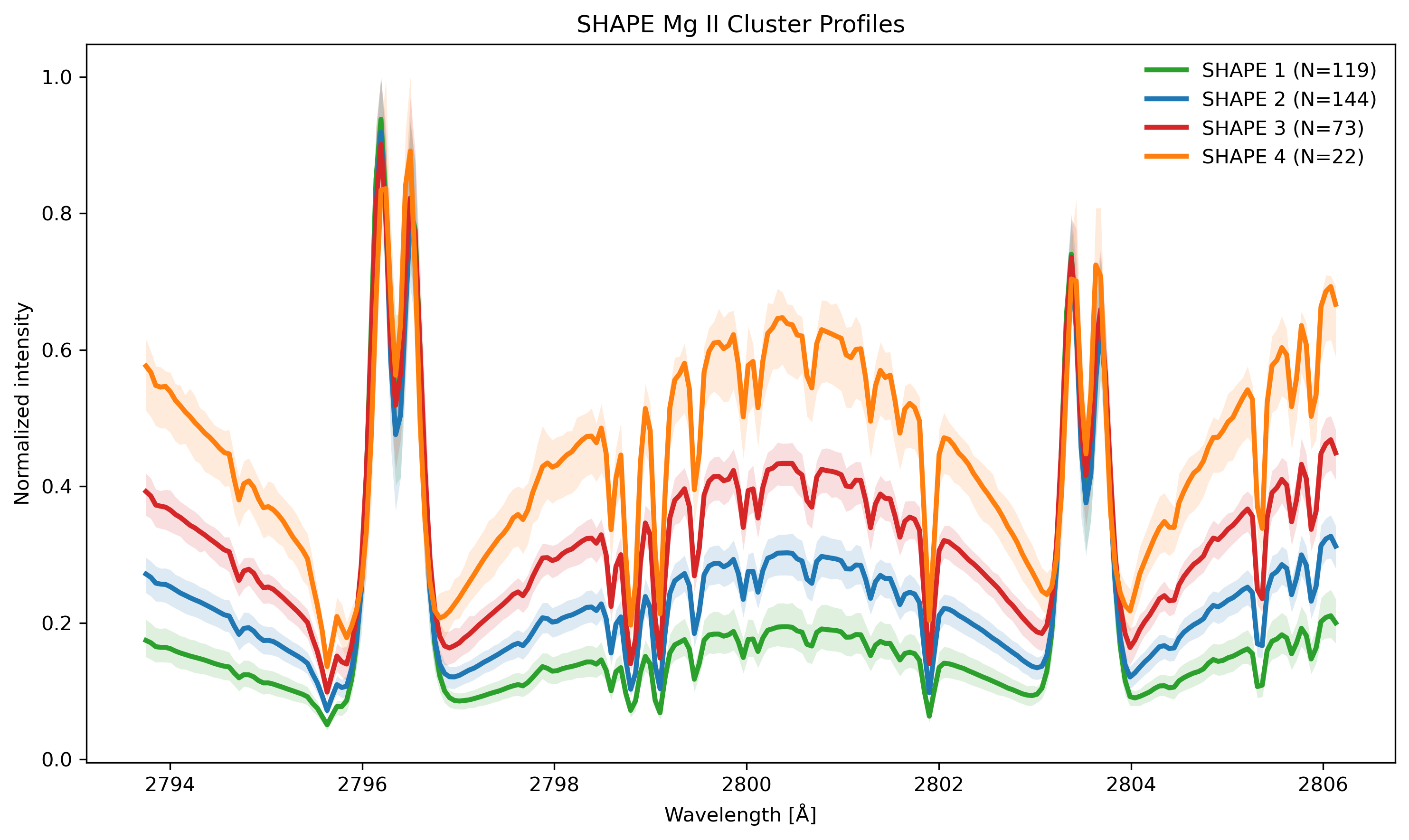}
\end{minipage}
\hfill
\begin{minipage}{0.48\textwidth}
\centering
\includegraphics[width=\linewidth]{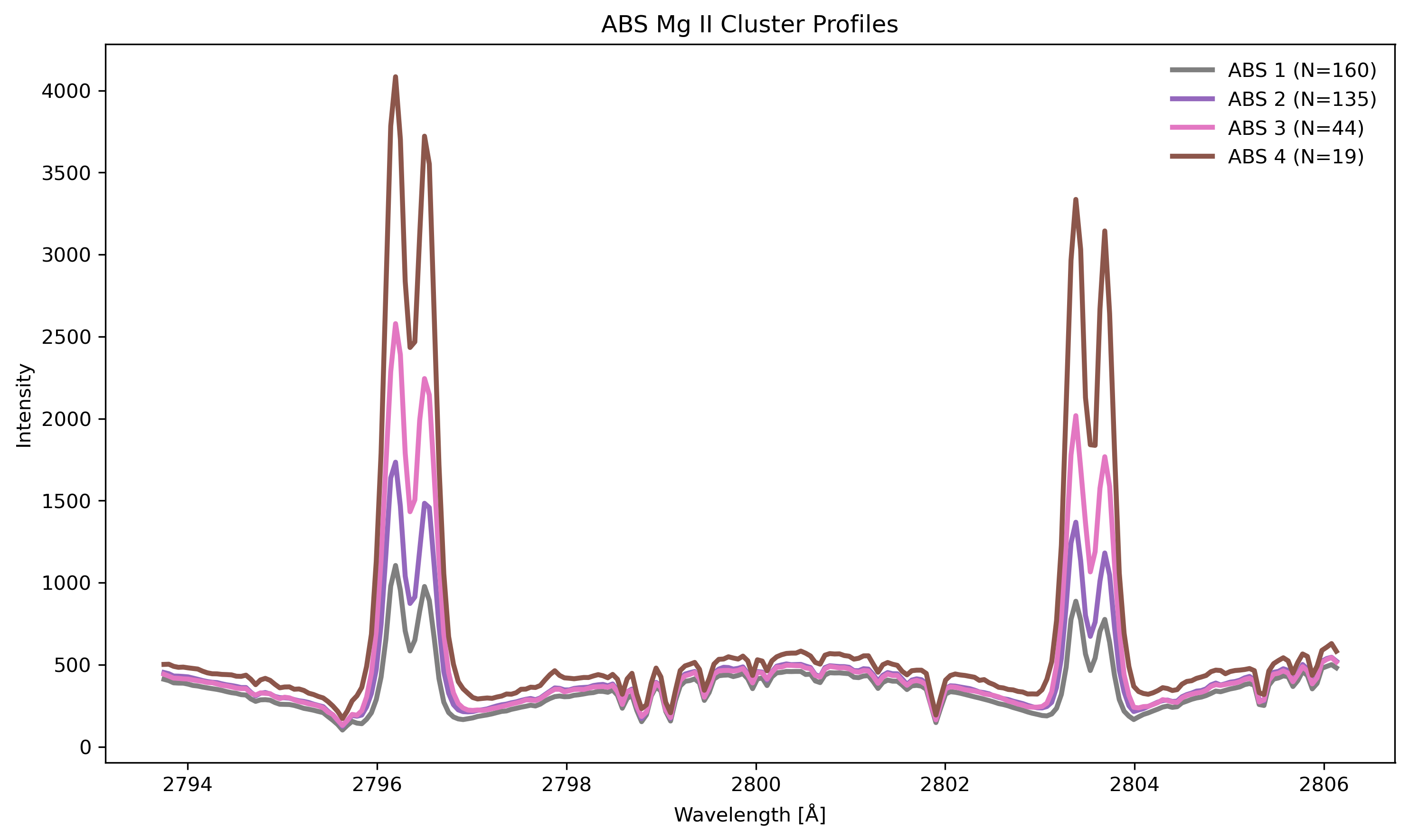}
\end{minipage}
\vspace{0.3cm}
\begin{minipage}{0.48\textwidth}
\centering
\includegraphics[width=\linewidth]{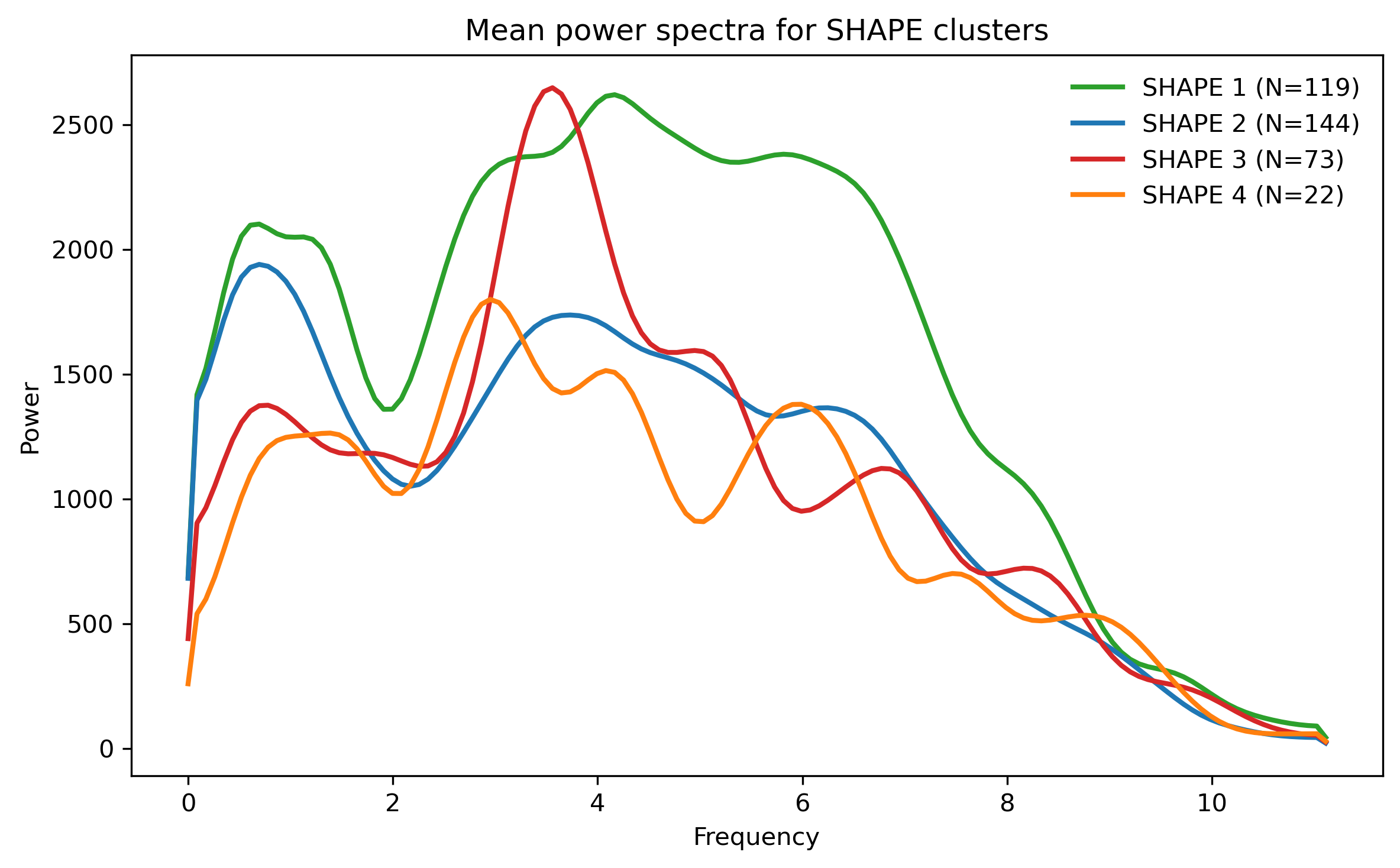}
\end{minipage}
\hfill
\begin{minipage}{0.48\textwidth}
\centering
\includegraphics[width=\linewidth]{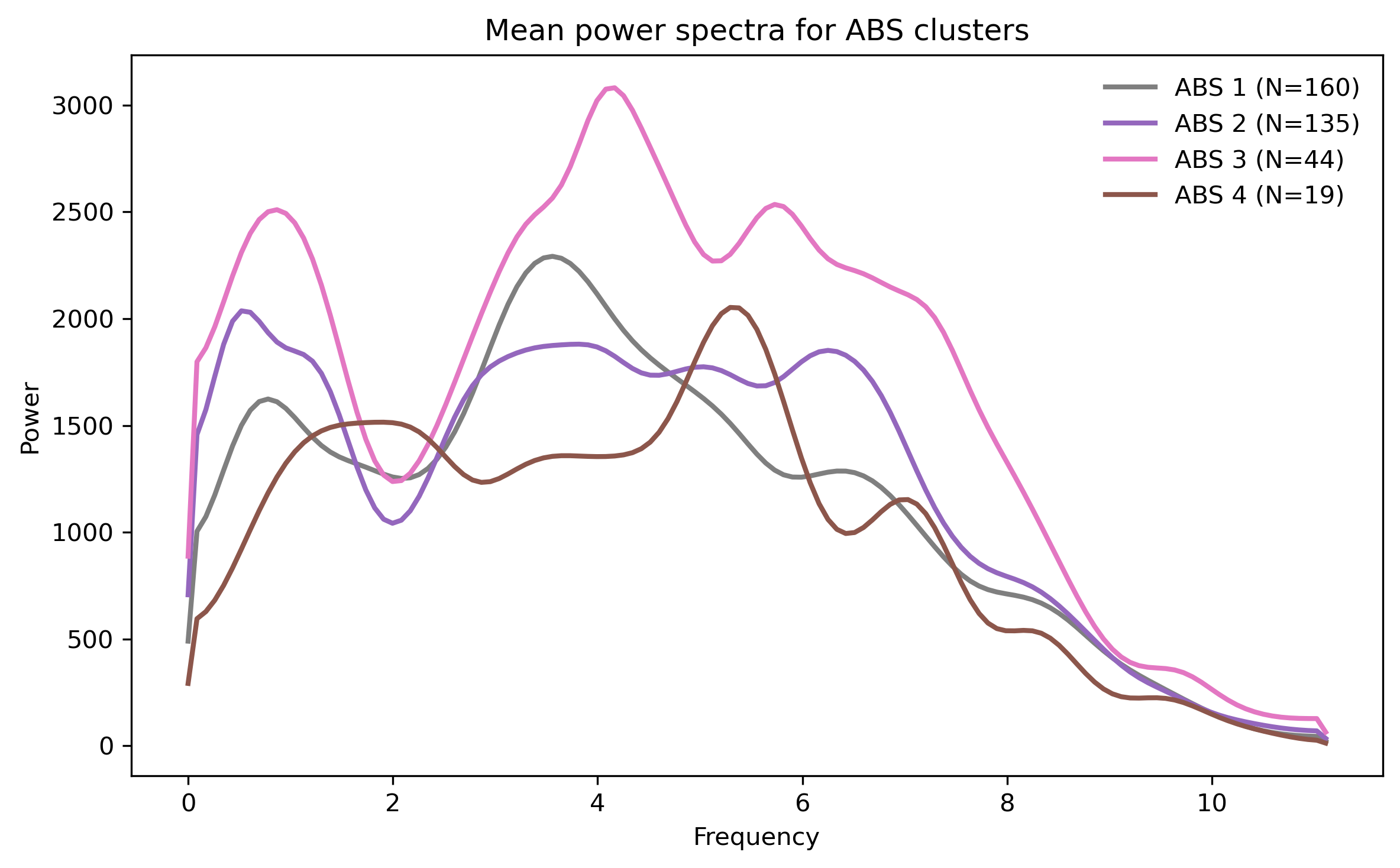}
\end{minipage}
\vspace{0.3cm}
\begin{minipage}{0.48\textwidth}
\centering
\includegraphics[width=\linewidth]{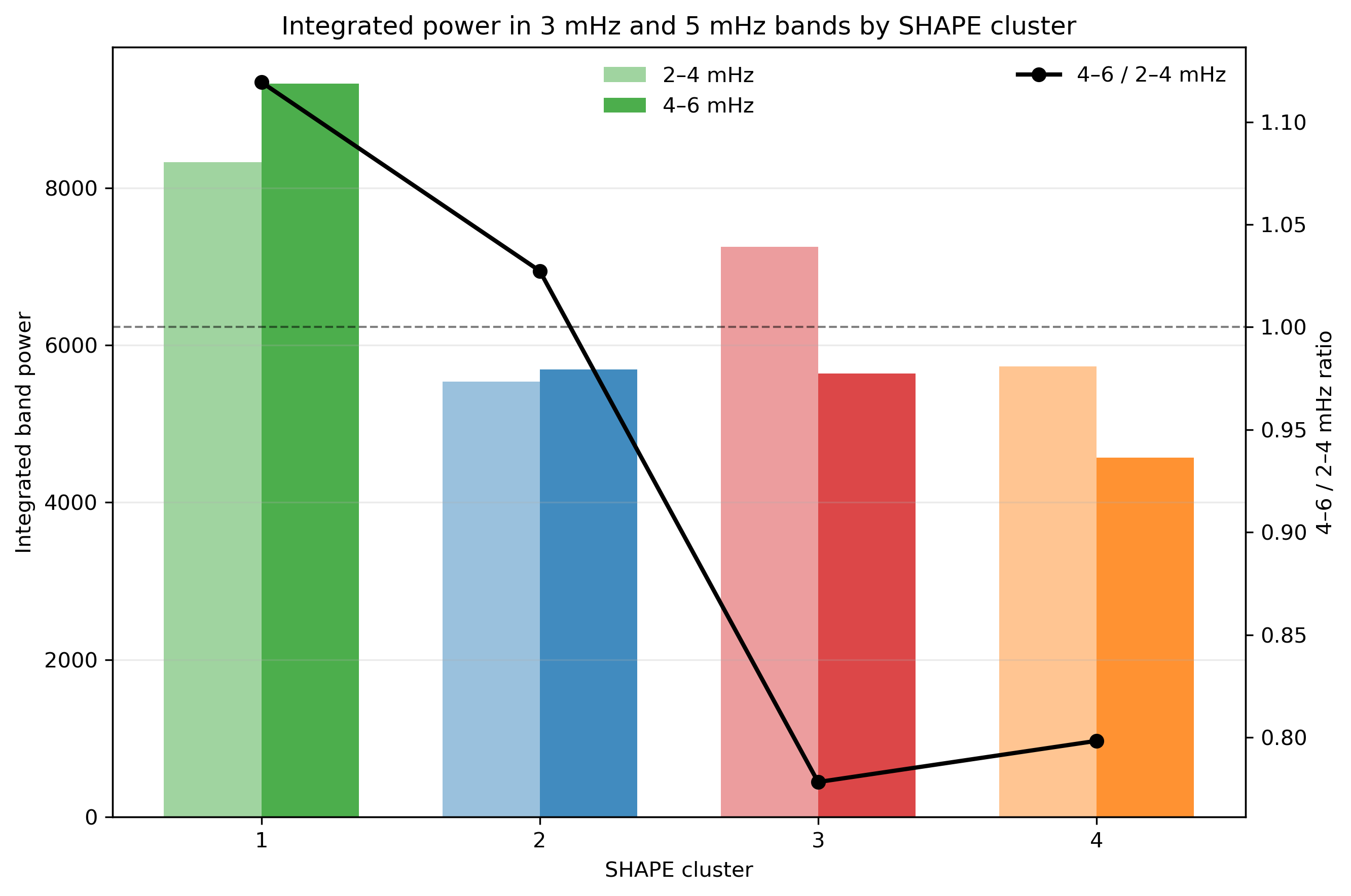}
\end{minipage}
\hfill
\begin{minipage}{0.48\textwidth}
\centering
\includegraphics[width=\linewidth]{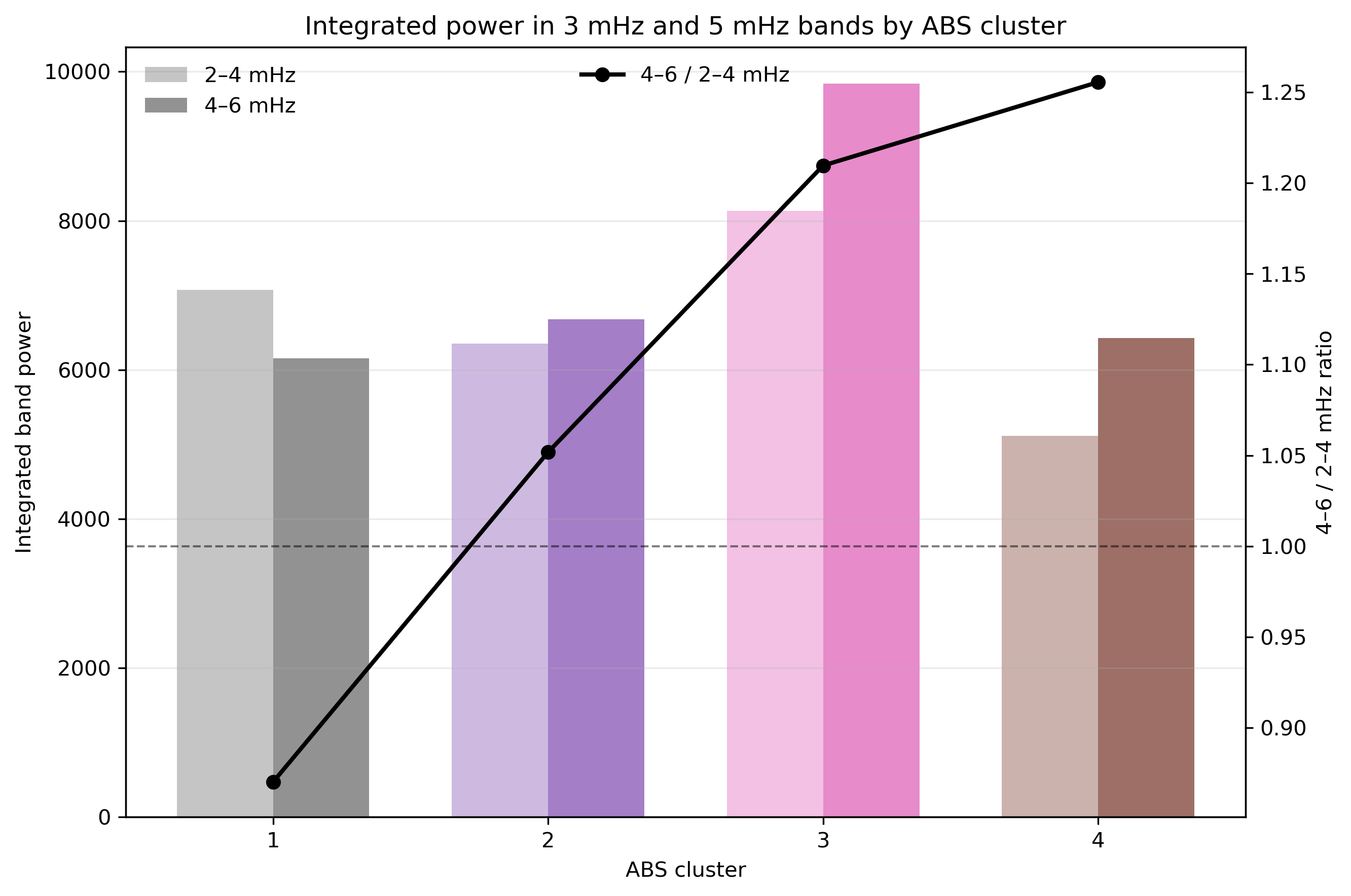}
\end{minipage}
\caption{
Comparison of \ion{Mg}{2} $h+k$ spectral morphology and wave-power properties for the SHAPE and ABS magnetic-element clusters. The left column shows results for the SHAPE clustering, while the right column shows results for the ABS clustering.
Top row: Median \ion{Mg}{2} $h+k$ line profiles for each cluster. 
Shaded regions of the SHAPE profiles indicate the interquartile range (25th–75th percentiles) of the cluster members.
Middle row: Mean power spectra of the horizontal velocities of photospheric magnetic elements derived from HMI magnetograms, averaged over the elements belonging to each cluster. 
Bottom row: Integrated oscillation power within the 2–4 mHz and 4–6 mHz bands for each cluster. Bars show the band-integrated power, while the black curve indicates the ratio of power in the 4–6 mHz band to that in the 2–4 mHz band. The dashed horizontal line marks equal power in the two bands (ratio = 1). 
}
\label{fig:fig2}
\end{figure}

\section{Chromospheric Spectral Classification}
\label{sec:kmeans}
\subsection{Construction of Element-Level \ion{Mg}{2} Profiles}
\label{sec:mg_profs}
Magnetic elements identified from the HMI magnetograms were represented by binary masks and reprojected into the IRIS coordinate system. 
Because each IRIS mosaic tile was acquired over approximately 34\,min, the reprojected masks were further corrected for the temporal scanning of the raster. 
A spatial shift based on the temporal offset between the HMI reference image and the observation time of each IRIS raster position was applied to every magnetic element. This correction aligns the magnetic-element masks with the effective observation time of the IRIS observations. The timing-corrected magnetic elements are shown in the lower right panel of Figure~\ref{fig:fig1}, where their contours are overplotted on the \ion{Mg}{2} $k$ integrated intensity mosaic. The contours trace the spatial distribution of the magnetic elements sampled across the AR--CH complex and provide the basis for the element-level spectral analysis described below.

An important consideration in associating the HMI magnetic elements with the IRIS spectra is that the two diagnostics sample different atmospheric heights. While HMI measures the photospheric magnetic field, the \ion{Mg}{2} spectra reflect conditions over a range of chromospheric heights. 
Magnetic fields may expand and become increasingly inclined with height, such that the chromospheric plasma observed by IRIS may be magnetically connected to photospheric locations displaced from the magnetic elements identified in the HMI observations. 
This introduces some uncertainty into the spatial association between the photospheric magnetic elements and their corresponding \ion{Mg}{2} spectral properties. The analysis therefore examines statistical relationships between photospheric magnetic elements and chromospheric conditions rather than assuming an exact one-to-one spatial correspondence between the two atmospheric layers.

The timing-corrected masks were used to identify all IRIS pixels associated with each magnetic element. Rather than clustering individual IRIS spectra, a representative \ion{Mg}{2} $h+k$ profile was constructed for each magnetic element. For a given element, all valid IRIS spectra contained within its contour were collected and the median spectrum was computed. Each magnetic element therefore contributed a single element-level \ion{Mg}{2} profile to the subsequent analysis, ensuring that the clustering reflects differences between magnetic structures rather than being dominated by elements containing large numbers of pixels.

The initial sample contained 495 magnetic elements distributed across the four IRIS mosaic tiles. 
To obtain reliable element-level statistics, only elements containing at least five valid IRIS spectra were retained. 
This criterion reduced the sample to 358 magnetic elements, which formed the final dataset used for the clustering analysis. 
The resulting dataset therefore consists of 358 representative \ion{Mg}{2} $h+k$ profiles, each corresponding to a single magnetic element.

These representative \ion{Mg}{2} $h+k$ spectra were processed following the procedures described in \citet{Bobra2019}. 
For each IRIS mosaic tile, the spectra were linearly interpolated onto a common wavelength grid spanning 2793.75--2806.13\,\AA, capturing the full \ion{Mg}{2} $k$ and $h$ line profiles while excluding neighboring spectral regions. 
The spectra were sampled using 244 wavelength bins, corresponding to the native IRIS spectral sampling across this wavelength range. 
A cleaning pipeline was applied independently to each tile to remove spectra affected by instrumental artefacts or unphysical values. 
Profiles containing intensity values $\leq$ --100 DN were excluded as invalid, while profiles with three or more wavelength bins at the maximum intensity were rejected as potentially saturated. 
After cleaning, the retained spectra from all four tiles were pooled for the clustering analysis.

\subsection{SHAPE and ABS Clustering Methodology}\label{sec:kemans_meth}

Two complementary clustering schemes were employed to isolate different aspects of the \ion{Mg}{2} spectral variability. 
In the first scheme, referred to as \textit{SHAPE} clustering, each spectrum was normalized prior to clustering so that the classification is sensitive primarily to profile morphology rather than absolute intensity. 
This approach emphasizes differences in spectral structure, including the depth of the central reversal, the relative strengths of the emission peaks, and variations in wing emission. 
In the second scheme, referred to as \textit{ABS} (absolute) clustering, the original cleaned spectra were used without normalization, preserving differences in overall emission strength. 
Comparison of the SHAPE and ABS classifications therefore distinguishes variations driven primarily by spectral morphology from those associated with emission strength.

The classification was performed using the MiniBatch $k$-means algorithm \citep{Lloyd1982} as implemented in the \texttt{scikit-learn} Python library \citep{Pedregosa2011}, adopting the default $k$-means++ centroid initialization. 
The algorithm partitions the spectra into $k$ groups by minimizing the variance within each cluster and determines a representative centroid profile for each group.

The number of clusters was evaluated using the elbow method and the silhouette score. For both SHAPE and ABS clustering, the largest improvement in cluster compactness and separation occurred for $k \approx 3$--$4$, with diminishing returns for larger values of $k$. 
We therefore adopt $k=4$ for both clustering schemes, providing a balance between capturing the diversity of the observed spectral profiles and maintaining statistically robust cluster populations.
Figure \ref{fig:fig2} (upper panels) shows the four SHAPE and ABS cluster profiles.

\begin{table}[!t]
\centering
\caption{Summary properties of the SHAPE and ABS cluster populations. The $k$-reversal depth fraction, $k$-line asymmetry, and $h+k$ core-to-wing ratio are defined in the Appendix. Magnetic fluxes are reported in units of $10^{17}$\,Mx.}
\label{tab:cluster_properties}
\renewcommand{\arraystretch}{1.15}

\begin{tabular*}{0.90\textwidth}{@{\extracolsep{\fill}}cccc@{}}
\toprule
\multicolumn{4}{c}{\textbf{SHAPE Spectral Morphology}} \\
\midrule
Cluster &
Reversal Depth Fraction ($D_k$) &
Asymmetry ($A_k$) &
Core-to-Wing Ratio ($R_{\rm CW}$) \\
\midrule
1 & 0.41 & 0.130 & 5.1 \\
2 & 0.45 & 0.125 & 3.3 \\
3 & 0.40 & 0.078 & 2.3 \\
4 & 0.36 & $-0.055$ & 1.6 \\
\midrule
\multicolumn{4}{c}{\textbf{Magnetic Properties}} \\
\midrule
Cluster &
Population &
Median $|B|$ (G) &
Median Flux ($10^{17}$\,Mx) \\
\midrule
1 & SHAPE (N = 119) & 31.5 & 9.9 \\
2 & SHAPE (N = 144) & 27.1 & 5.7 \\
3 & SHAPE (N = 73) & 21.0 & 3.4 \\
4 & SHAPE (N = 22) & 22.9 & 3.4 \\
\midrule
1 & ABS (N = 160) & 23.6 & 4.7 \\
2 & ABS (N = 135) & 29.6 & 6.7 \\
3 & ABS (N = 44) & 30.8 & 12.0 \\
4 & ABS (N = 19) & 37.8 & 15.0 \\
\bottomrule
\end{tabular*}
\end{table}

\subsection{Quantitative Characterization of SHAPE and ABS Classes}

Table~\ref{tab:cluster_properties} summarizes the spectral diagnostics and magnetic properties associated with each SHAPE and ABS class (see Appendix \ref{app:defs} for formal definitions of all metrics). 
The spectral diagnostics are measured from the cluster centroid profiles and therefore characterize the representative spectral morphology of each class rather than the distribution of the individual member spectra. 
The representative profiles are shown in the upper panels of Figure~\ref{fig:fig2}, with the shaded interquartile ranges for the SHAPE classes illustrating the variability among cluster members.
Conversely, the magnetic properties are computed from all magnetic elements assigned to each cluster and are reported as the median properties of the cluster population, thereby characterizing the typical magnetic environment associated with each spectral class. 
Together, these quantities provide a compact description of the systematic differences underpinning the classification.

To facilitate physical interpretation, it is useful to consider the diagnostic properties of the spectral metrics used in the classification. 
The core-to-wing ratio, $R_{\mathrm{CW}}$, provides a measure of the relative contribution of chromospheric line-core emission compared to photospheric or lower-chromospheric wing intensity, and is therefore primarily sensitive to the level of chromospheric heating and emissivity. 
In comparison, the reversal depth fraction, $D_k$, characterizes the strength of the central depression in the line core and reflects the vertical structure of the source function and opacity in the upper chromosphere (see Sections 5--7 of \cite{Leenaarts2013b} and Section 10 of \cite{Leenaarts2013a}). 
As such, it is sensitive to non-LTE radiative transfer effects, including the decoupling of the source function from local thermodynamic conditions, rather than serving as a direct proxy for heating alone.
The asymmetry parameter, $A_k$, captures differences between the blue and red emission peaks and is particularly sensitive to velocity gradients and flows in the line formation region. 
These diagnostics provide complementary constraints on the thermodynamic structure and dynamical state of the chromosphere, enabling a more comprehensive interpretation of the observed spectral variations.

Across the SHAPE classes, clear trends are evident in all spectral metrics. The core-to-wing ratio, $R_{\mathrm{CW}}$, decreases markedly from 5.1 in SHAPE~1 to 1.6 in SHAPE~4, indicating a progressive reduction in core emission relative to the wings. 
The reversal depth fraction, $D_k$, shows a corresponding decline from 0.45 (SHAPE~2) to 0.36 (SHAPE~4), consistent with a transition toward less strongly self-reversed profiles.
The asymmetry parameter, $A_k$, evolves systematically across the sequence: SHAPE~1--3 exhibit positive asymmetries (0.130, 0.125, and 0.078), while SHAPE~4 shows a negative value ($-0.055$). This sign change reflects a reversal in the relative dominance of the blue and red emission peaks.

The magnetic properties associated with the SHAPE classes show structured, though non-monotonic, variation. 
The median unsigned field strength, $|B|$, decreases from 31.5~G in SHAPE~1 to 21.0~G in SHAPE~3, before increasing slightly to 22.9~G in SHAPE~4. 
A similar behavior is seen in the median magnetic flux, which declines from $9.9 \times 10^{17}$~Mx in SHAPE~1 to $3.4 \times 10^{17}$~Mx in SHAPE~3 and SHAPE~4. 
These trends indicate that the SHAPE sequence is broadly associated with decreasing magnetic concentration, although the most extreme spectral morphology (SHAPE~4) does not correspond to the weakest-field regime, possibly due to the low number of magnetic elements.

In contrast, the ABS classes exhibit a clear monotonic ordering in magnetic properties. 
Both the median field strength and magnetic flux increase from ABS~1 to ABS~4, with $|B|$ rising from 23.6~G to 37.8~G and the median flux increasing from $4.7 \times 10^{17}$~Mx to $1.5 \times 10^{18}$~Mx. This progression indicates that the ABS classification spans a range from relatively weak-field to strongly magnetized environments.

The metrics presented in Table~\ref{tab:cluster_properties} demonstrate systematic variations in spectral morphology across the SHAPE classes and in magnetic properties across both SHAPE and ABS classes.
The consistency of these trends supports the robustness of the classification scheme and provides a quantitative basis for the physical interpretation of the classes.
\section{Oscillatory Properties of the Spectral Classes}\label{sec:kmeans_osc}

\subsection{Mean Power Spectra}\label{sec:spec}

Figure \ref{fig:fig2} (middle row) shows the mean power spectra of the horizontal velocities of photospheric magnetic elements for the SHAPE and ABS clusters. 
These spectra were obtained by averaging the individual magnetic element velocity power spectra within each cluster, providing a representative distribution of oscillatory power as a function of frequency.

To assess the reliability of the observed oscillatory features, we performed a randomization analysis, following
an approach similar to that of \cite{Jess2023} and \cite{Berretti2024}.
The horizontal velocity time series were randomly permuted
1000 times, and the resulting power spectra were used to
derive frequency dependent confidence thresholds for the
ensemble average horizontal power in each SHAPE and ABS
class (details are included in Appendix~\ref{app:stats}).
This process destroys any real temporal oscillations or periodicities while preserving the overall distribution of velocity values.
Several prominent spectral peaks exceed the 95th percentile
threshold, particularly in the 4--7\,mHz range.
These results indicate that the observed frequency enhancements are not readily reproduced by temporally randomized velocity data, providing confidence that the prominent oscillatory features are not dominated by random fluctuations in the velocity measurements.

The SHAPE clusters exhibit systematic differences in both the amplitude and frequency distribution of oscillatory power. 
All four clusters display broad power enhancements spanning the 0--8\,mHz frequency range. 
SHAPE~1 generally exhibits the highest power levels and maintains substantial power out to frequencies approaching 9\,mHz. 
SHAPE~2 displays a similarly broad spectral form, although with lower overall power. 
SHAPE~3 exhibits a more pronounced concentration of power near 3\,mHz, while SHAPE~4 displays multiple peaks of power extending from $\sim$3\,mHz into the high-frequency tail. 
These differences indicate that magnetic elements associated with different chromospheric spectral morphologies exhibit distinct oscillatory signatures.

The ABS clusters likewise display clear differences in their mean power spectra. 
ABS~3 exhibits the highest power levels across much of the measured frequency range and retains substantial power to frequencies approaching 9\,mHz, with a broad maximum near 3--4\,mHz. 
ABS~2 displays a similar spectral form, although with lower power levels and a less extended high-frequency tail. 
ABS~1 exhibits a more localized enhancement near 3\,mHz, whereas ABS~4 displays a distinct power enhancement in the 5--6\,mHz range. 
The differences between the ABS spectra extend across a broad range of frequencies rather than being confined to a single spectral feature.

The mean power spectra reveal systematic differences between both the SHAPE and ABS classifications. 
These differences are evident in both the overall power levels and the distribution of power across frequency. 
To quantify these trends, the following section examines the integrated power within the 2--4\,mHz and 4--6\,mHz frequency bands and the relative balance between them.

\subsection{Frequency-Band Power Analysis}\label{sec:power}

Building on these spectral differences, the oscillatory power was integrated over two frequency ranges: 2--4\,mHz and 4--6\,mHz. 
These bands were selected to sample both the dominant photospheric $p$-mode oscillations and the higher frequency oscillations commonly observed in the chromosphere. 
Figure \ref{fig:fig2} (bottom row) shows the integrated power in each band together with the ratio of the 4--6\,mHz power to the 2--4\,mHz power for the SHAPE and ABS classifications. 
This allows us to identify a possible high frequency enhancement in the spectra.  

The SHAPE clusters exhibit substantial differences in both the integrated power and the relative distribution of power between the two frequency bands. 
SHAPE~1 displays the largest integrated power in both bands and is the only cluster for which the 4--6\,mHz power clearly exceeds the 2--4\,mHz power, resulting in a ratio of 1.12. 
SHAPE~2 exhibits nearly equal power in the two bands, with a ratio close to unity (1.03), whereas SHAPE~3 and SHAPE~4 exhibit greater power in the 2--4\,mHz band than in the 4--6\,mHz band, producing ratios of 0.78 and 0.80, respectively. 
These results indicate a transition from spectra with enhanced higher-frequency power (SHAPE~1) to those in which lower-frequency oscillations contribute a larger fraction of the total power (SHAPE~3 and SHAPE~4).

The ABS clusters exhibit a different pattern. 
ABS~3 displays the largest integrated power in both frequency bands, consistent with its strong broadband power spectrum, whereas ABS~4 exhibits lower total power despite having the largest band-power ratio. 
The ratio increases systematically from ABS~1 (0.87) to ABS~4 (1.26), passing through ABS~2 (1.05) and ABS~3 (1.21). 
This progression indicates a gradual shift toward relatively stronger high-frequency oscillations across the ABS sequence. 
Thus, although the ABS clusters are primarily distinguished by emission strength, they also exhibit systematic differences in the distribution of oscillatory power between lower- and higher-frequency bands.

The band-power analysis demonstrates that the SHAPE and ABS classifications are associated with systematic differences in oscillatory behavior. 
Both classification schemes exhibit clear variations in the balance between lower- and higher-frequency oscillations. 
For the SHAPE clusters, this balance changes systematically with spectral morphology, whereas for the ABS clusters it varies systematically across the emission-strength sequence. 

\section{Implications for Wave Propagation and Energy Conversion}

The SHAPE and ABS classifications can be interpreted in the context of kink mode wave dynamics and their coupling to the chromospheric plasma. 
The measured oscillatory power spectra correspond primarily to transverse motions, indicating that kink mode waves provide a significant component of the observed wave energy \citep{Stangalini2013}. 
Although kink waves are often treated as nearly incompressible in the thin-flux-tube approximation \citep{Spruit1981,Edwin1983}, they become weakly compressible under realistic chromospheric conditions through stratification, flux-tube expansion, nonlinear effects, and mode coupling \citep[e.g.][]{Wedemeyer2009,Jess2015,Stangalini2015}. 
Consequently, the \ion{Mg}{2} spectral diagnostics do not always respond directly to the transverse displacements themselves, but rather to the associated thermodynamic and velocity perturbations of the plasma. 
The SHAPE and ABS trends therefore reflect not only the presence of kink waves, but the extent to which their energy is converted into compressive motions and contributes to enhanced chromospheric emission within different magnetic environments. 
In this interpretation, the association between the SHAPE and ABS classes and different balances of $p$-mode and higher frequency oscillatory power indicates that the chromosphere responds differently to the propagation and dissipation of kink wave energy, producing the distinct \ion{Mg}{2} spectral profiles observed in each class.

A key aspect in interpreting the SHAPE and ABS classifications is the range of chromospheric heights sampled by the \ion{Mg}{2} diagnostics relative to the propagation and dissipation of kink mode wave energy. 
Observational constraints from multiple studies suggest that kink waves propagate nonlinearly from the photosphere into the lower chromosphere with relatively little dissipation. 
In particular, \citet{Stangalini2015} report enhanced chromospheric power in magnetic elements across frequencies above $\sim$2.6\,mHz, with power amplification in the 1--8\,mHz range derived from \ion{Ca}{2} diagnostics. 
Given that the \ion{Ca}{2} line core forms at heights of order $\sim$400–500\,km, this implies that a substantial fraction of kink wave energy remains largely undepleted through the photosphere and lower chromosphere. 
At the same time, studies of wave transmission into higher layers indicate that significant damping occurs before energy reaches the corona. 
For example, \citet{Morton2014} show that chromospheric kink wave power spectra differ from their coronal counterparts, with evidence for frequency-dependent attenuation consistent with enhanced dissipation in the upper chromosphere and lower corona, likely associated with gradients across the transition region. 

These results collectively suggest that the \ion{Mg}{2} lines, which sample an extended height range spanning the upper-photosphere to upper chromosphere \citep[e.g.][]{Leenaarts2013a,Leenaarts2013b,Pereira2013,DePontieu2021}, probe a regime where kink wave energy is progressively redistributed and dissipated. 
Within this framework, the observed variations in line shape and emission are consistent with a scenario in which the \ion{Mg}{2} diagnostics capture the transition from predominantly undamped wave propagation at lower heights to increased mode coupling, compressive response, and heating at greater heights \citep{DePontieu2021}.

The dependence of both SHAPE and ABS classifications on magnetic field strength and flux suggests that stronger magnetic elements act as more efficient waveguides \citep[e.g.][]{Jess2015}. 
In these structures, kink waves can propagate with reduced attenuation, not strongly constrained by the acoustic cutoff that limits compressive motions \citep{Spruit1981,Morton2014,Jess2015}, allowing a greater fraction of wave energy to reach chromospheric heights. 
This enhanced energy flux increases the likelihood of both dissipation and mode coupling. 
In particular, partial conversion of transverse wave energy into compressive perturbations is expected in regions where the plasma $\beta$ approaches unity \citep{Bogdan2003}. 
Such processes can generate velocity gradients and localized compressions that contribute to the observed increases in line asymmetry and higher core-to-wing ratios, while also influencing the central reversal depth through changes in the vertical stratification and source function structure \citep{Leenaarts2013a,Leenaarts2013b} associated with SHAPE~1 and 2.
At the same time, dissipative processes, including resonant absorption and phase mixing, provide plausible pathways for converting transverse wave energy into heat \citep{Vandoor2020}, consistent with the monotonic increase in emission observed across the ABS sequence.

The frequency dependence of the power spectra further constrains this interpretation. 
Although kink waves are not subject to a strict acoustic cutoff and can propagate across both the $p$-mode (2--4\,mHz)  and high-frequency (4--6\,mHz) bands, the efficiency with which their energy is transferred to the chromospheric plasma is expected to depend on frequency. 
Higher-frequency oscillations are more readily able to generate small scale spatial gradients in velocity and thermodynamic structure \citep{Vandoor2020} and are therefore more susceptible to dissipative processes and mode coupling. 
This is consistent with the enhanced high-frequency power associated with stronger magnetic elements and with the more pronounced chromospheric signatures observed in SHAPE~1 and 2 and ABS~3 and 4.
In contrast, lower-frequency oscillations near 3\,mHz, while capable of transporting energy, are likely to be less efficient at driving compressive responses and heating, particularly in weaker magnetic elements where waveguiding is less effective. 
This provides a natural explanation for the reduced emission levels and more photospheric-like spectral profiles observed in SHAPE~3–4 and ABS~1–2.

It should be noted, however, that compressive slow magnetoacoustic waves are also expected to contribute to the observed spectral characteristics. 
In comparison to kink modes, these waves are sensitive to the acoustic cutoff and therefore cannot freely propagate unless this cutoff is reduced. 
In regions where the effective acoustic cutoff is lowered, for example by magnetic field inclination or non-adiabatic effects, slow-mode waves may propagate into the chromosphere and directly generate shocks and velocity gradients \citep[e.g.][]{Depontieu2004,Centeno2006}. 
The spectral signatures associated with SHAPE~1 and 2 are therefore likely to arise from a combination of processes, including both direct compressive wave propagation and the indirect effects of kink wave energy conversion. 
Disentangling these contributions is not straightforward, and the present analysis does not uniquely distinguish between them.

The combined SHAPE and ABS behavior can be interpreted as evidence that magnetic field strength and structure regulate both the transmission of wave energy and the efficiency with which that energy is converted into observable chromospheric signatures. 
Strong-field, high-flux magnetic elements support efficient propagation of kink mode waves and provide favorable conditions for their coupling and dissipation, leading to enhanced chromospheric emission and dynamic spectral structure. 
Weaker magnetic elements limit both the transmission and conversion of wave energy, resulting in lower emission levels and more weakly structured line profiles. 
Within this framework, the observed \ion{Mg}{2} spectral properties and frequency-dependent power distributions provide a diagnostic of how kink wave dominated energy transport is processed within the chromosphere, while also allowing for a non-negligible contribution from compressive wave modes.

\section{Conclusions}

We have examined the relationship between chromospheric \ion{Mg}{2} spectral properties and photospheric kink mode oscillations in magnetic elements within an active region–coronal hole complex. 
By combining element-resolved power spectra with unsupervised clustering of \ion{Mg}{2} $h+k$ line profiles, we identify systematic links between wave dynamics, magnetic structure, and chromospheric emission.

The key findings are as follows:

\begin{enumerate}

\item The SHAPE clustering shows that distinct chromospheric spectral morphologies are associated with different oscillatory signatures. 
The classes exhibiting stronger line asymmetries and larger core-to-wing ratios, indicative of enhanced chromospheric emission, preferentially show enhanced high-frequency power, with SHAPE~1 displaying a clear dominance of the $4$--$6$\,mHz band over the $p$-mode band.

\item The ABS clustering demonstrates a monotonic increase in the relative importance of high-frequency oscillations with increasing emission strength and magnetic flux. This indicates that stronger magnetic elements preferentially support the transmission or persistence of higher-frequency wave power.

\item These results show that high-frequency oscillations are not uniformly distributed, but instead depend on both chromospheric state and magnetic environment. The SHAPE classification isolates the spectral conditions under which high-frequency power is most dominant, whereas the ABS classification captures the broader trend of increasing high-frequency dominance with increasing magnetic field strength, unsigned flux, and chromospheric emission.

\item Our findings are consistent with a scenario in which kink mode waves transport energy into the chromosphere, where part of this energy is converted into compressive motions and contributes to enhanced chromospheric emission through mode coupling and dissipative processes. 
This conversion appears to be most effective in structured and strongly magnetized environments, giving rise to the observed spectral variability.

\end{enumerate}

\begin{acks}
We thank Julio Hernandez Camero for valuable assistance in $k$-means clustering analysis. 
IRIS is a NASA small explorer mission developed and operated by the Lockheed Martin Solar and Astrophysics Laboratory (LMSAL) with mission operations executed at NASA Ames Research Center and major contributions to downlink communications funded by the Norwegian Space Centre (NSC, Norway) through an ESA PRODEX contract.  
D.B. acknowledge support from UKSA1478
grant No. UKRI920.
A.S.H.T. acknowledges funding from the UPSaclaySTAR-Phy programme under the Marie Skłodowska-Curie Actions (MSCA COFUND), Grant Agreement No. 101216532.
S.L.Y. is grateful to the Science Technology and Facilities Council for the award of an Ernest Rutherford Fellowship (ST/X003787/1).
MB acknowledges that this publication was produced while attending the PhD programme in Space Science and Technology at the University of Trento, Cycle XXXIX, with the support of a scholarship financed by the Ministerial Decree no. 118 of 2nd March 2023, based on the NRRP - funded by the European Union - NextGenerationEU - Mission 4 ``Education and Research'', Component 1 ``Enhancement of the offer of educational services: from nurseries to universities'' - Investment 4.1 ``Extension of the number of research doctorates and innovative doctorates for public administration and cultural heritage'' - CUP E66E23000110001.
\end{acks}
\appendix
\section{Definitions of Mg II k Spectral Profile Metrics}\label{app:defs}
\begin{figure}[!t]
\centering
\includegraphics[width=0.90\linewidth]{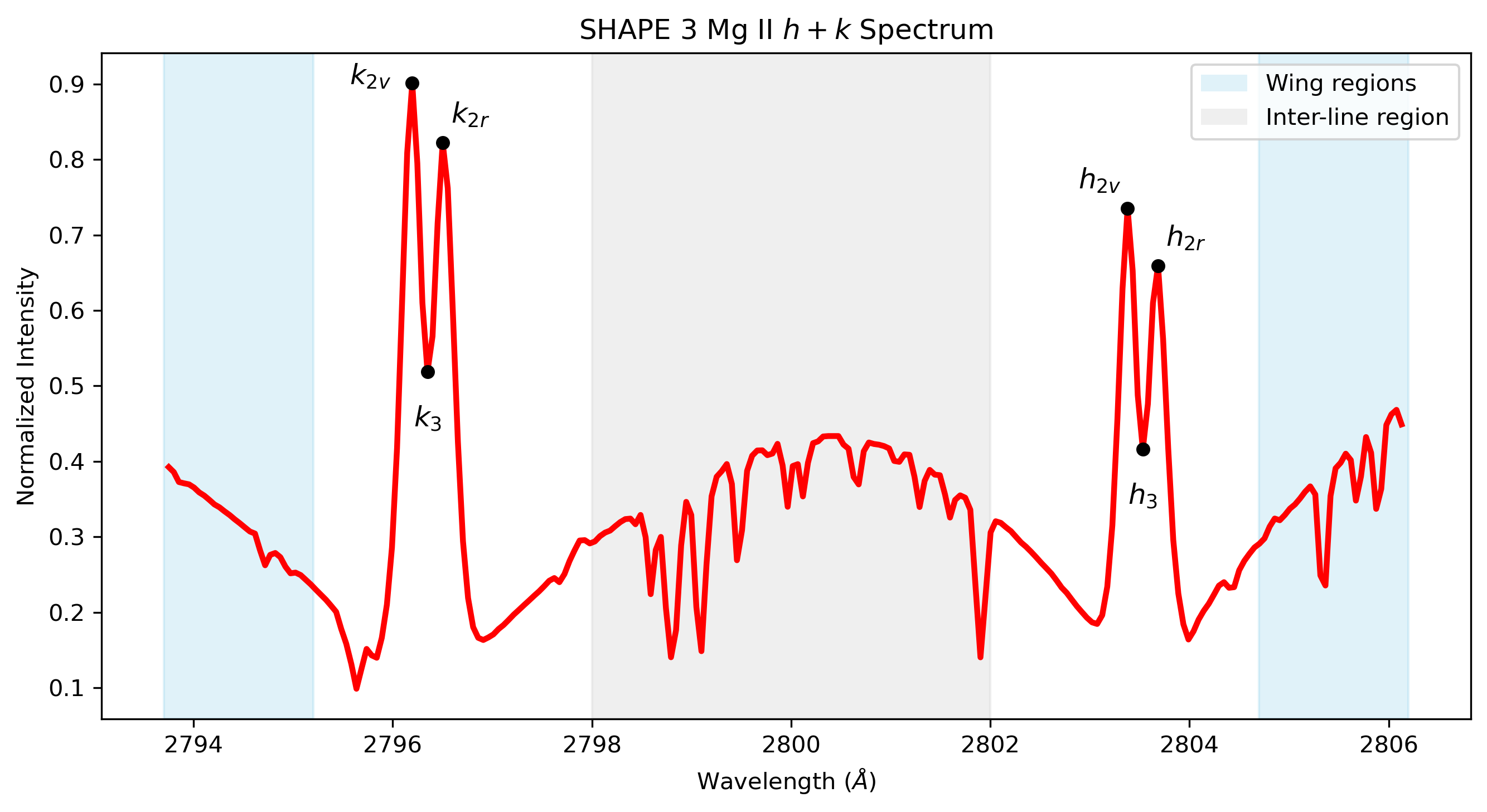}
\caption{
Example SHAPE~3 \ion{Mg}{2} $h+k$ spectrum used to illustrate the spectral morphology diagnostics. 
Filled circles mark the standard \ion{Mg}{2} spectral features $k_{2v}$, $k_3$, $k_{2r}$, $h_{2v}$, $h_3$, and $h_{2r}$. 
The shaded blue regions indicate the wing intervals, and the shaded gray region indicates the inter-line interval used in the denominator of the core-to-wing ratio. 
The numerator of the ratio is calculated from the mean intensities of the four emission peaks, $k_{2v}$, $k_{2r}$, $h_{2v}$, and $h_{2r}$.
}
\label{fig:appendix_spectrum}
\end{figure}

\begin{figure*}[!t]
\centering
\includegraphics[width=0.48\textwidth]{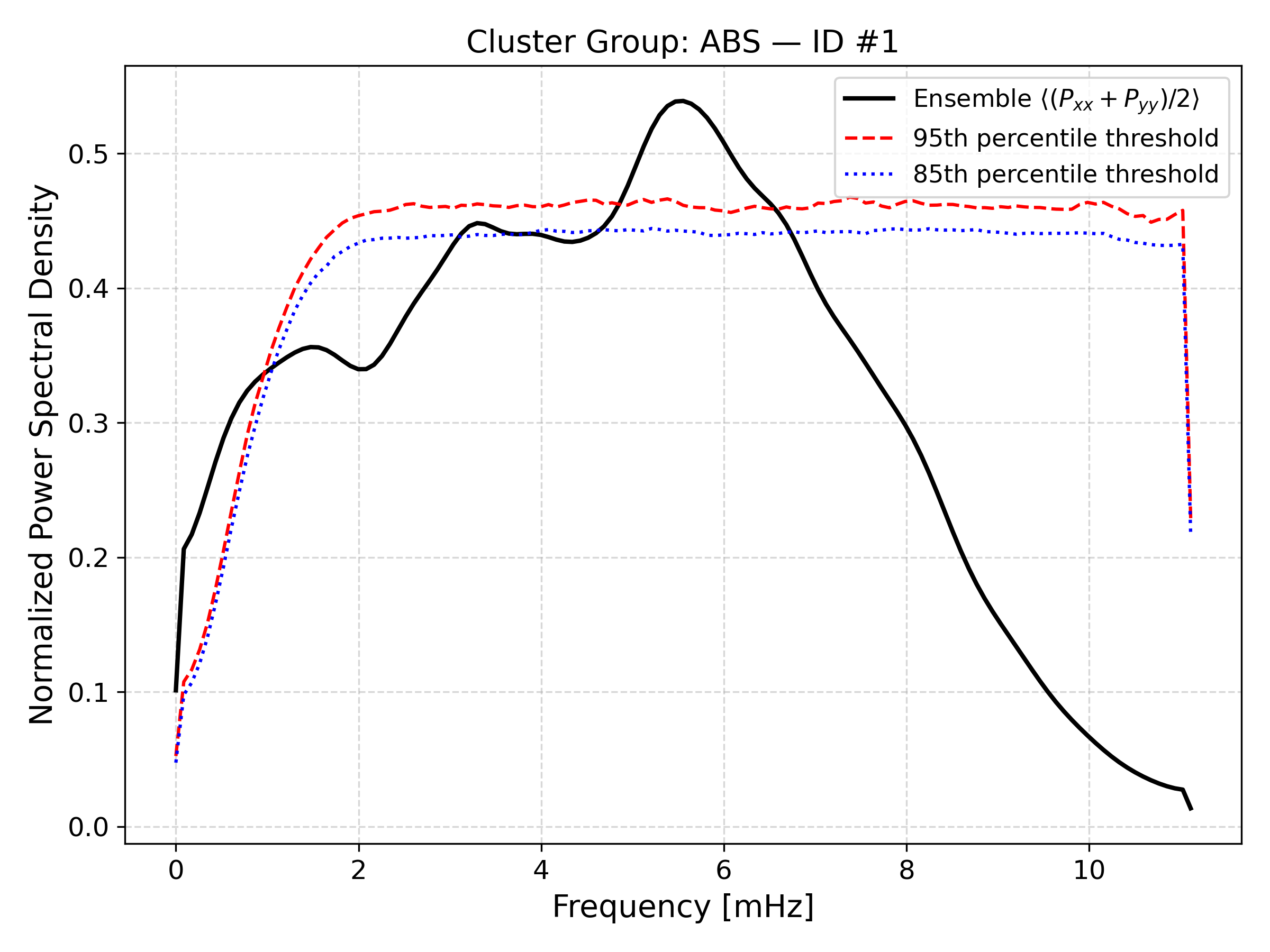}
\includegraphics[width=0.48\textwidth]{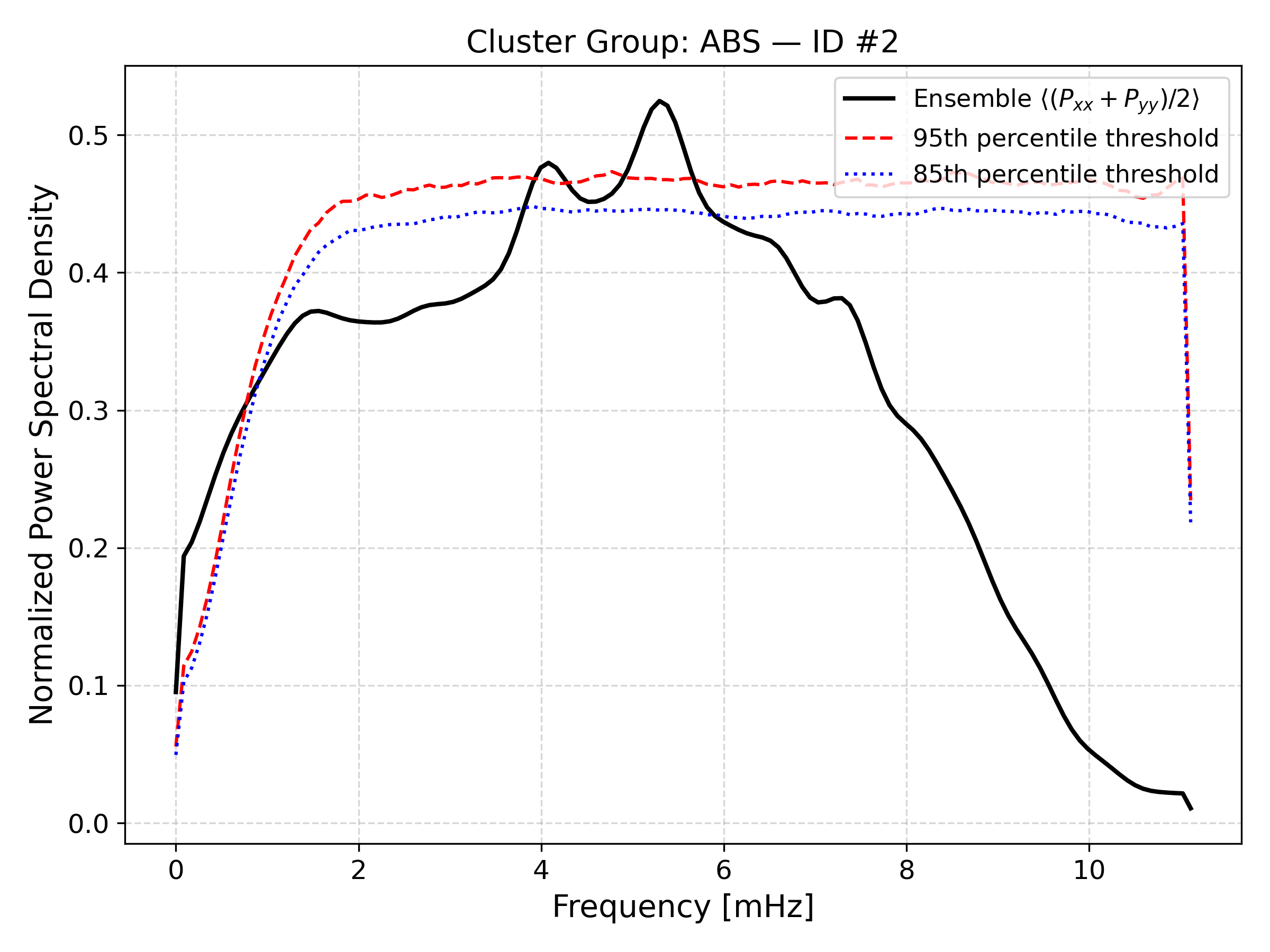}

\includegraphics[width=0.48\textwidth]{
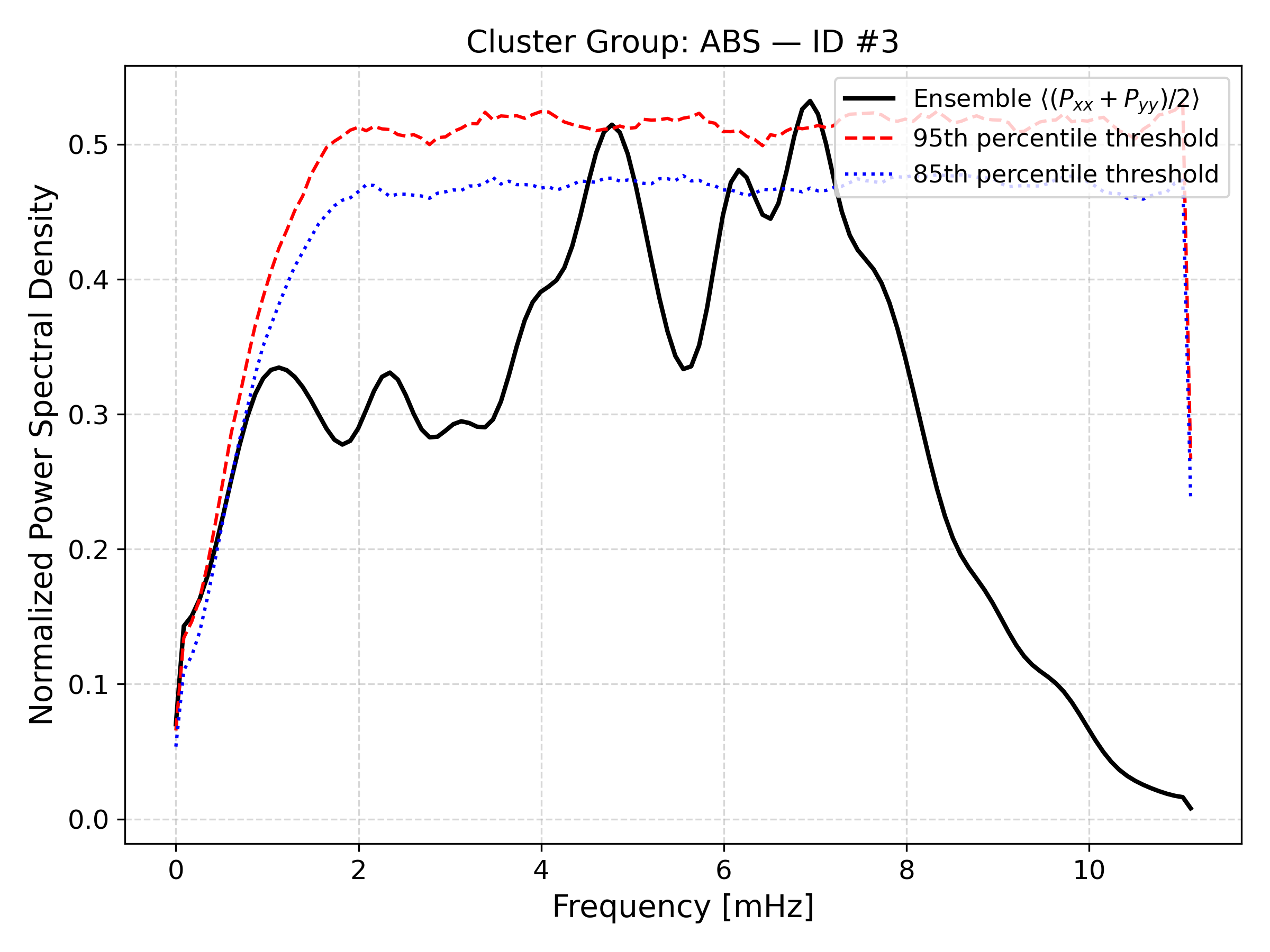}
\includegraphics[width=0.48\textwidth]{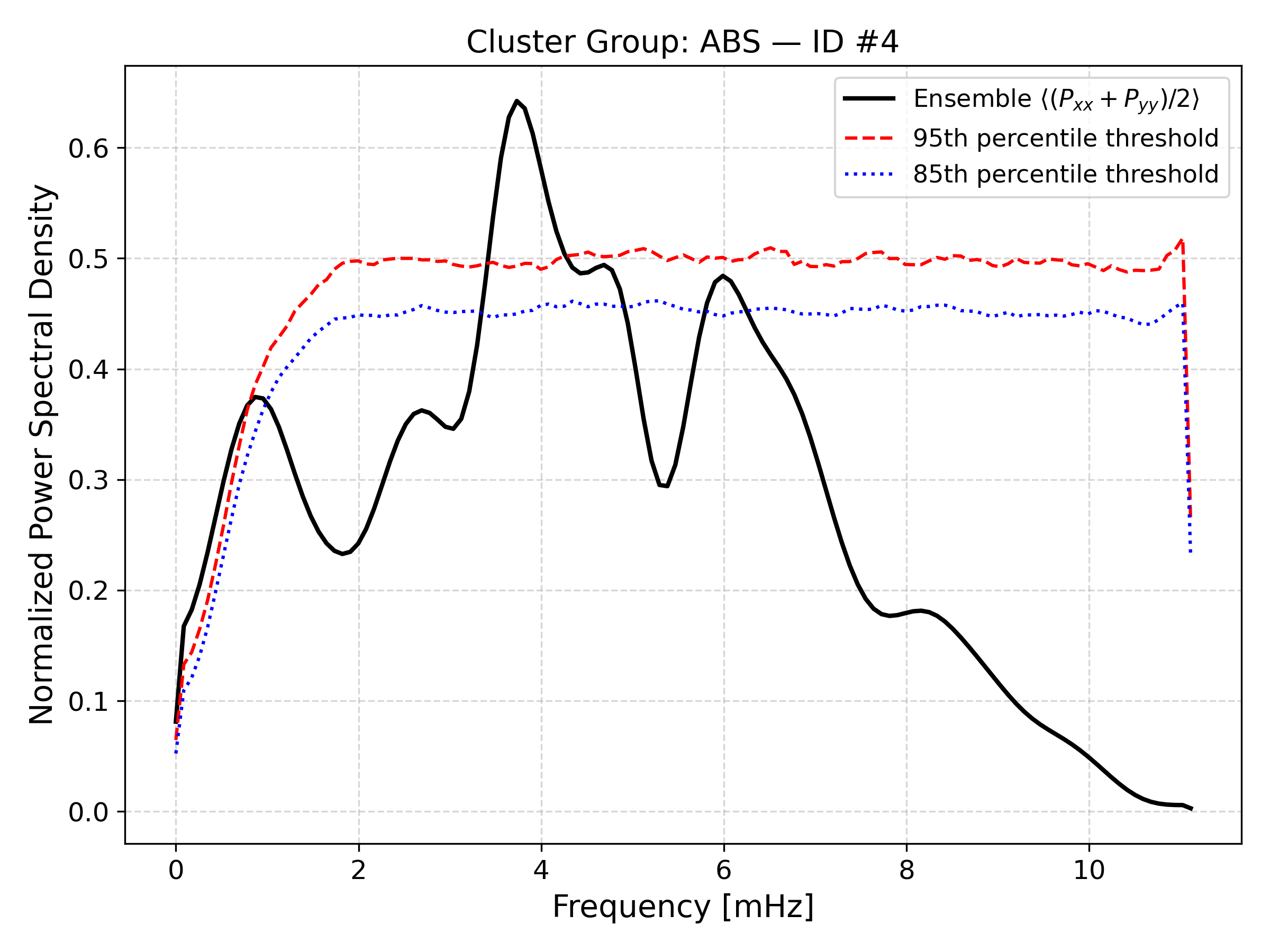}

\includegraphics[width=0.48\textwidth]{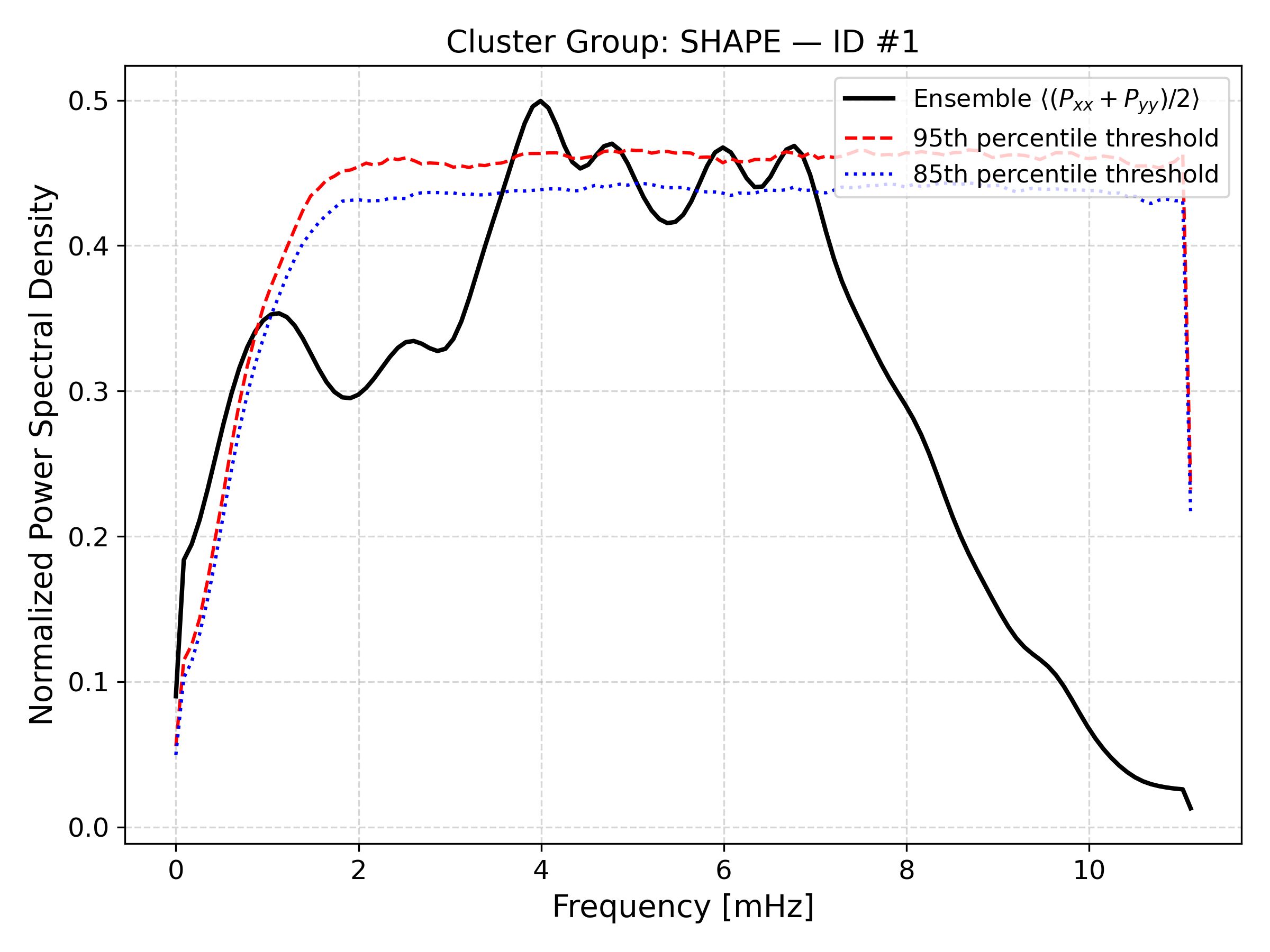}
\includegraphics[width=0.48\textwidth]{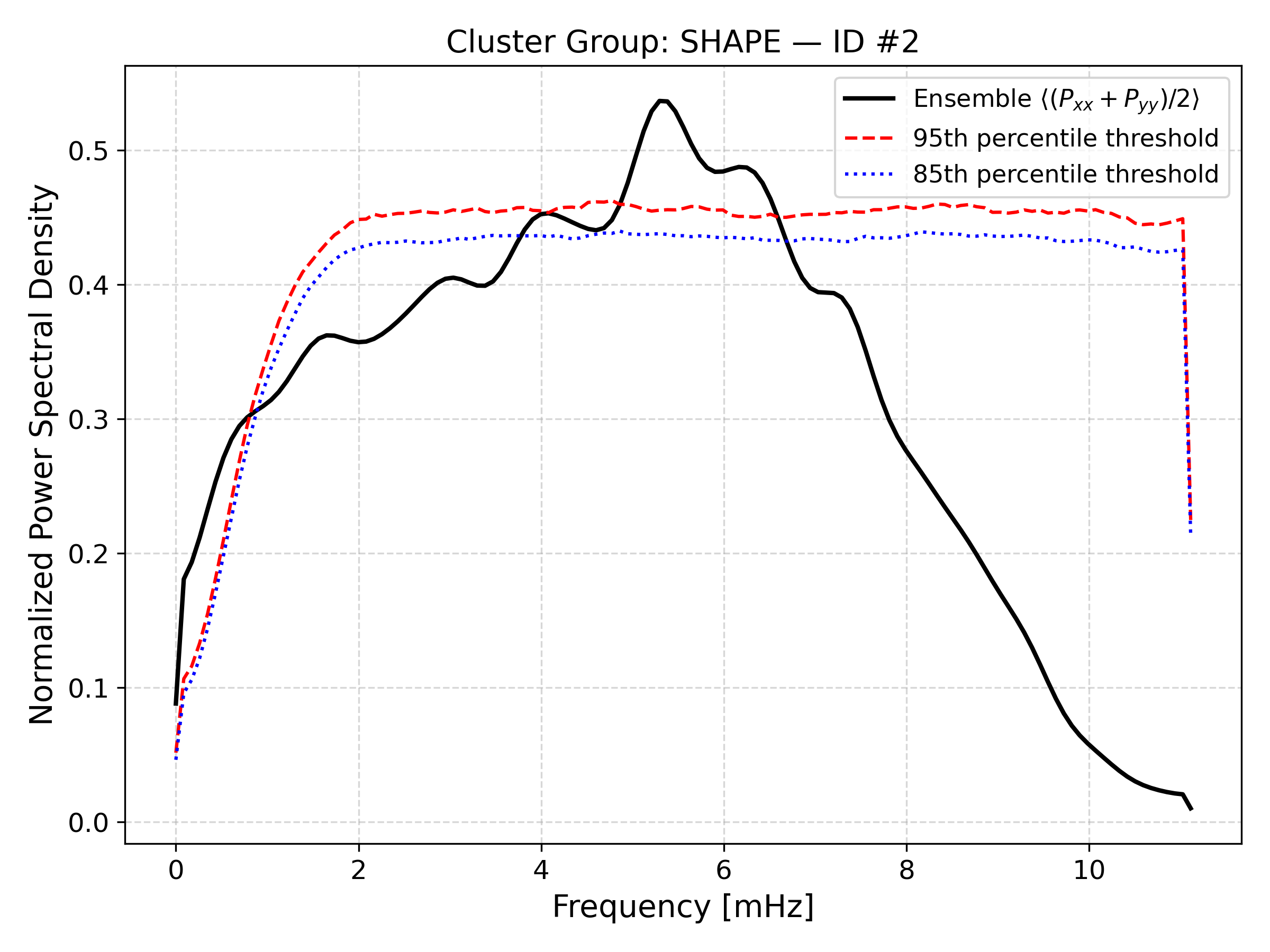}

\includegraphics[width=0.48\textwidth]{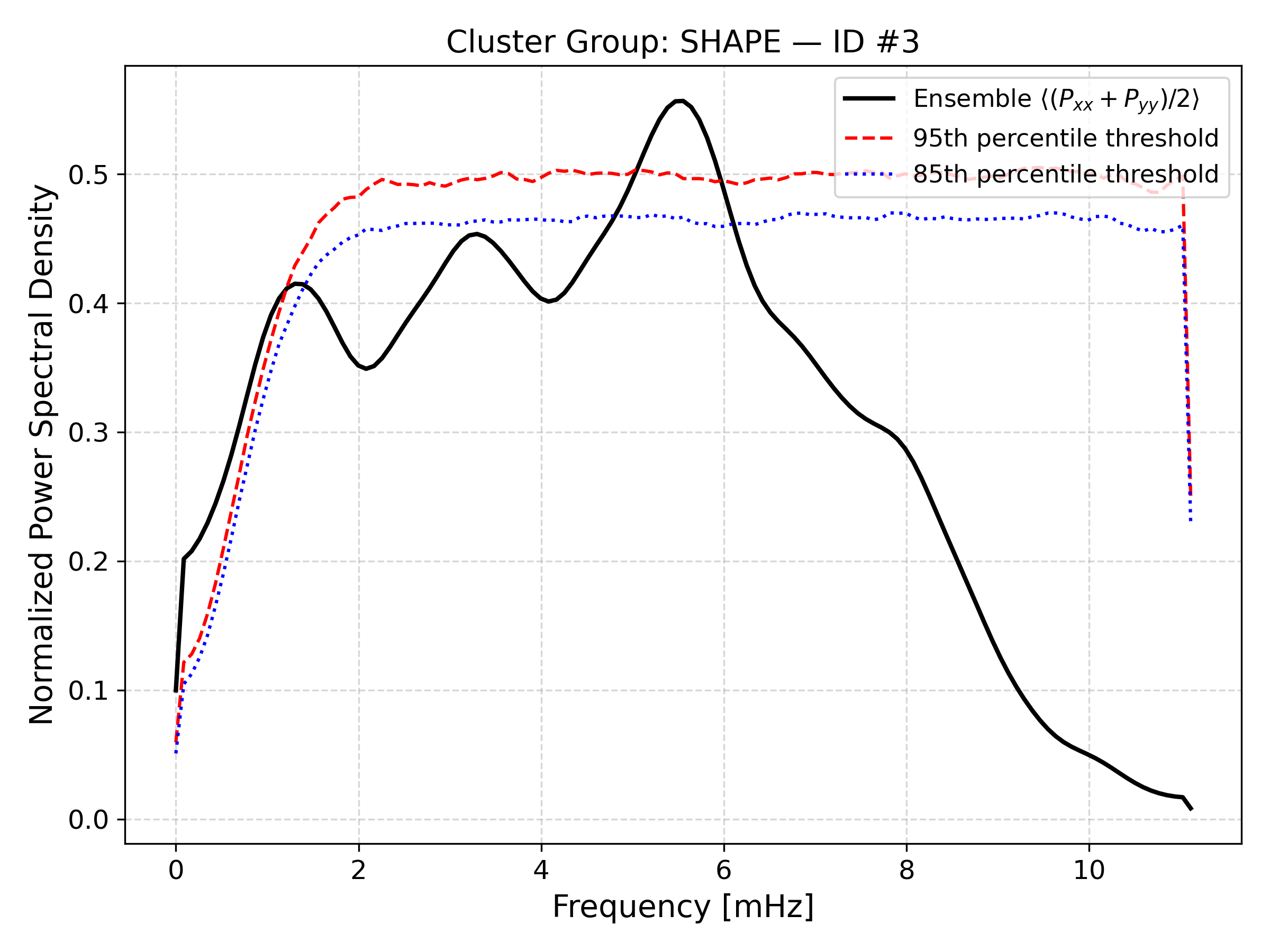}
\includegraphics[width=0.48\textwidth]{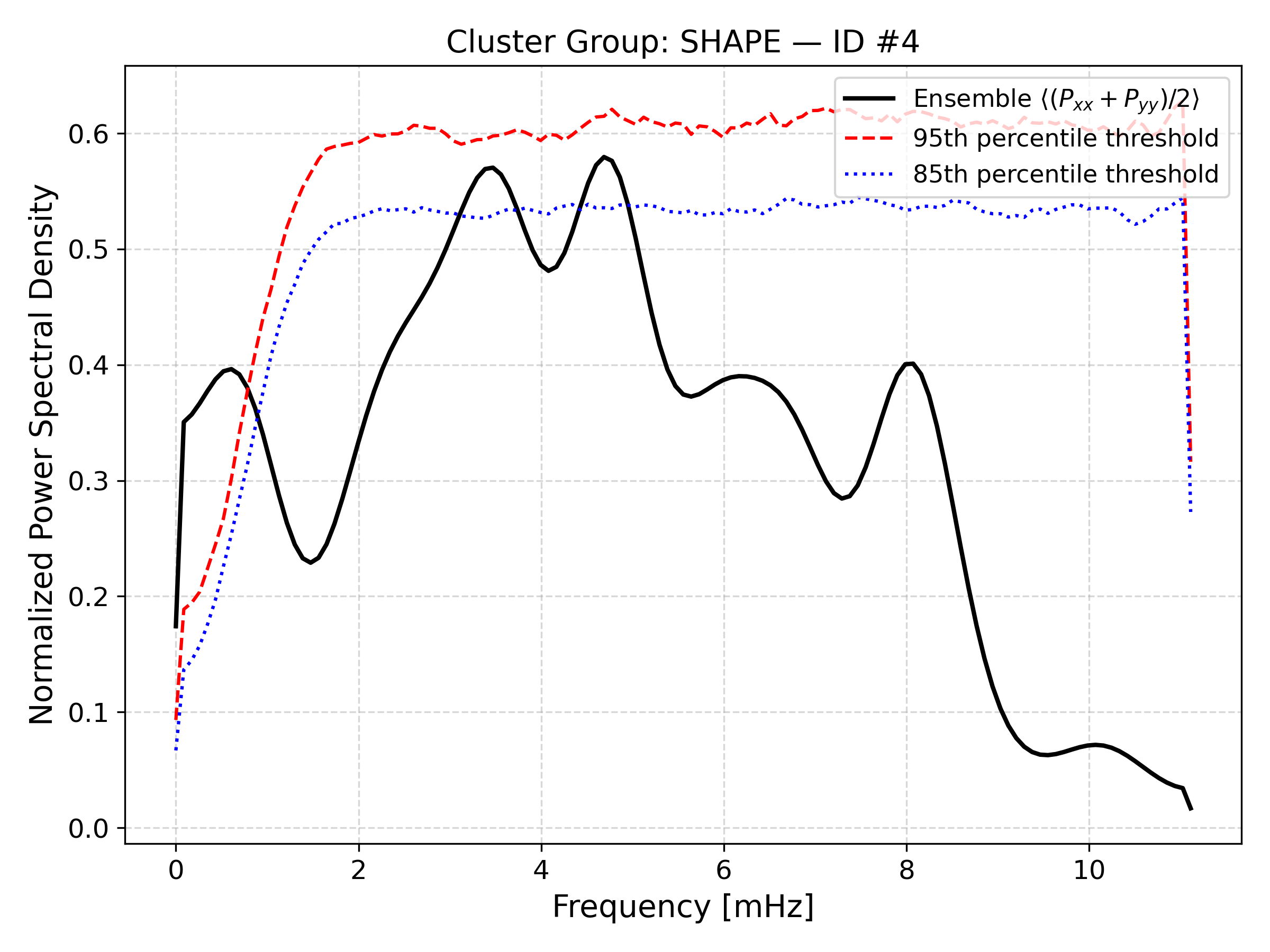}

\caption{Ensemble-average horizontal power spectra for the four ABS classes (top two rows) and four SHAPE classes (bottom two rows). Black curves show the observed normalized ensemble-average spectra, while blue dotted and red dashed curves indicate the 85th and 95th percentile thresholds, respectively, derived from 1000 temporal randomizations.}
\label{fig:stats_osc}
\end{figure*}

To quantify the spectral morphology of the SHAPE clusters, we calculated three diagnostics from the cluster centroid \ion{Mg}{2} $h+k$ profiles: the $k$-line reversal depth fraction, the $k$-line asymmetry, and the $h+k$ core-to-wing ratio. 
Figure~\ref{fig:appendix_spectrum} illustrates the spectral features and wavelength intervals used to construct the three morphology diagnostics. The labeled $h_{2v}$, $h_3$, $h_{2r}$, $k_{2v}$, $k_3$, and $k_{2r}$ features define the emission peaks and central reversals, while the shaded wing and inter-line regions indicate the wavelength intervals used in the core-to-wing ratio.
Because the SHAPE clustering was performed using normalized profiles, these diagnostics characterize the relative morphology of the Mg II profiles rather than their absolute emission strength.

For both the \ion{Mg}{2} $k$ and $h$ lines, the blue and red emission peaks ($k_{2v}$, $k_{2r}$, $h_{2v}$, and $h_{2r}$) were identified on either side of the line center. 
The central reversal ($k_3$ or $h_3$) was then defined as the minimum intensity located between the two emission peaks (Figure~\ref{fig:appendix_spectrum}).
The mean peak intensity was defined as

\begin{equation}
I_{\rm peak}
=
\frac{I(k_{2v})+I(k_{2r})}{2},
\end{equation}
where $I(k_{2v})$ and $I(k_{2r})$ denote the intensities of the blue and red emission peaks, respectively.
The reversal depth fraction reported in Table~\ref{tab:cluster_properties} was calculated for the \ion{Mg}{2} $k$ line as

\begin{equation}
D_k
=
\frac{I_{\rm peak}-I(k_3)}
     {I_{\rm peak}}.
\end{equation}
This diagnostic addresses the question: \emph{How deep is the central reversal relative to the emission peaks?}
Larger values correspond to deeper central reversals and therefore more strongly self-reversed line profiles. Because the central reversal forms higher in the chromosphere than the surrounding emission peaks, this metric provides a quantitative measure of the strength of the self-reversal while remaining independent of the overall brightness of the profile.

The $k$-line asymmetry was calculated from the relative strengths of the red and blue emission peaks,

\begin{equation}
A_k
=
\frac{I(k_{2r})-I(k_{2v})}
     {I(k_{2r})+I(k_{2v})}.
\end{equation}
This diagnostic addresses the question: \emph{Which emission peak is stronger?}
Positive values indicate profiles with stronger red peaks ($k_{2r}>k_{2v}$), negative values indicate profiles with stronger blue peaks ($k_{2v}>k_{2r}$), and values near zero correspond to approximately symmetric peak intensities. Peak asymmetries are commonly interpreted as signatures of velocity gradients and differential plasma motions along the line of sight within the chromosphere.

The core-to-wing ratio was calculated using the mean of the $h$ and $k$ emission-peak intensities as a measure of line-core emission. 
The wing intensity was defined as the mean intensity across three wavelength intervals: the blue wing (2793.70--2795.20~\AA), the inter-line region (2798.00--2802.00~\AA), and the red wing (2804.70--2806.20~\AA). The core-to-wing ratio was then defined as

\begin{equation}
R_{\rm CW}
=
\frac{\langle I_{\rm peak,k}, I_{\rm peak,h}\rangle}
     {\langle I_{\rm blue\ wing},
              I_{\rm interline},
              I_{\rm red\ wing}\rangle}.
\end{equation}
This diagnostic addresses the question: \emph{How strong is the chromospheric emission relative to the surrounding lower-atmospheric emission?}
Larger values of $R_{\rm CW}$ indicate stronger chromospheric emission peaks relative to the surrounding wing and inter-line emission, whereas smaller values indicate a progressively larger contribution from the lower chromosphere and upper photosphere. Because the reversal depth is quantified separately by $D_k$, the core-to-wing ratio primarily measures the relative prominence of the chromospheric emission peaks rather than the depth of the central reversal itself.

These diagnostics were calculated only for the SHAPE clusters because SHAPE clustering isolates differences in normalized spectral morphology. They are not reported for the ABS clusters because the ABS classification was performed on unnormalized profiles and primarily separates profiles according to overall emission level rather than detailed line shape.

\section{Statistical Assessment of Oscillatory Power}\label{app:stats}

The randomization analysis was carried out separately for
each of the four SHAPE and four ABS classes. 
For the magnetic elements, the temporal ordering of the horizontal velocity components, $V_x$ and $V_y$, were independently randomized 1000 times. 
Power spectra were then recalculated using the same procedure applied to the observed velocity time series.

For each randomization, the horizontal power spectra were
combined as $(P_{xx}+P_{yy})/2$ and averaged over the
magnetic elements belonging to each cluster across all
four mosaic tiles. 
The resulting 1000 surrogate
ensemble average spectra per class were processed and normalized consistently with the observed spectra.
The 85th and 95th percentiles of the surrogate distributions were then determined at each frequency,
providing frequency dependent reference thresholds.

Figure~\ref{fig:stats_osc} compares the observed spectra
with these thresholds for the ABS (top two rows) and
SHAPE (bottom two rows) classifications.
For the ABS classification, peaks meeting or exceeding the 95th percentile threshold are present in all four classes. 
Their frequencies and relative strengths vary between clusters, with most enhancements occurring between approximately 3 and 7\,mHz.
ABS 4 is notable for a pronounced peak near 3.7\,mHz.

For SHAPE 1, 2, and 3, prominent peaks exceed the 95th percentile
threshold, particularly in the approximately 4--6\,mHz range. 
In contrast, the peaks in SHAPE 4
remain below this threshold, but above the 85th percentile threshold, possibly due to the small number of magnetic elements in this class.

These comparisons identify the particular spectral features that exceed the randomization derived
thresholds. 
The thresholds are evaluated at individual frequencies and do not establish statistically
significant differences between the cluster populations.
\bibliographystyle{spr-mp-sola}
\bibliography{IRIS_bib}

\end{document}